\documentclass[twocolumn,,trackchanges]{aastex62}
\usepackage{amsmath}
\hypersetup{
	colorlinks	= true,
	linkcolor	= red,
	urlcolor	= cyan,
	citecolor	= blue
}

\usepackage{amssymb}

\usepackage{lipsum}
\usepackage{multirow}
\usepackage{rotating}
\usepackage{comment}

\graphicspath{{./images/}}

\newcommand{\iraclis}{\mbox{\textsc{iraclis}}}
\newcommand{\exot}{\mbox{\textsc{ExoTETHyS}}}
\newcommand{\taurex}{\mbox{\textsc{Tau-REx}}}
\newcommand{\exotransmit}{\mbox{\textsc{Exo-Transmit}}}
\newcommand{\multinest}{\mbox{\textsc{MultiNest} }}

\turnoffediting

\usepackage{fancyhdr}
\shortauthors{Damiano et al.}

\begin{document}
	
	\title{A transmission spectrum of the sub-Earth planet L98-59~b in 1.1 -- 1.7 $\mu$m}
	
	\correspondingauthor{Mario Damiano}
	\email{mario.damiano@jpl.nasa.gov}
	
	\author[0000-0002-1830-8260]{Mario Damiano}
	\affiliation{Jet Propulsion Laboratory, California Institute of Technology, Pasadena, CA 91109, USA}
	
	\author[0000-0003-2215-8485]{Renyu Hu}
	\affiliation{Jet Propulsion Laboratory, California Institute of Technology, Pasadena, CA 91109, USA}
	\affiliation{Division of Geological and Planetary Sciences, California Institute of Technology, Pasadena, CA 91125, USA}
	
    \author[0000-0001-7139-2724]{Thomas Barclay}
	\affiliation{University of Maryland, Baltimore County, 1000 Hilltop Cir, Baltimore, MD 21250, USA}
	\affiliation{NASA Goddard Space Flight Center, 8800 Greenbelt Rd, Greenbelt, MD 20771, USA}
	
	\author[0000-0003-0562-6750]{Sebastian Zieba}
	\affiliation{Max-Planck-Institut für Astronomie (MPIA), Königstuhl 17, 69117 Heidelberg, Germany}
	\affiliation{Leiden Observatory, Leiden University, Niels Bohrweg 2, 2333CA Leiden, The Netherlands}
	
	\author[0000-0003-0514-1147]{Laura Kreidberg}
	\affiliation{Max-Planck-Institut für Astronomie (MPIA), Königstuhl 17, 69117 Heidelberg, Germany}
	
	\author[0000-0002-2072-6541]{Jonathan Brande}
	\affiliation{Department of Physics and Astronomy, University of Kansas, 1082 Malott, 1251 Wescoe Hall Dr., Lawrence, KS 66045, USA}
	
	\author[0000-0001-8020-7121]{Knicole D. Colon}
	\affiliation{NASA Goddard Space Flight Center, 8800 Greenbelt Rd, Greenbelt, MD 20771, USA}
	\affiliation{NASA GSFC Sellers Exoplanet Environments Collaboration}
	
	\author[0000-0002-2553-096X]{Giovanni Covone}
	\affiliation{Dipartimento di Fisica ``E. Pancini", Università di Napoli Federico II, Via Cinthia I-80126 Naples, Italy}
	\affiliation{INAF, Osservatorio Astronomico di Capodimonte, Salita Moiariello, Napoli, Italy}
	\affiliation{INFN, Sezione di Napoli, C.U. Monte S. Angelo, Via Cinthia, I-80126 Napoli, Italy}
	
	\author{Ian Crossfield}
	\affiliation{Department of Physics and Astronomy, University of Kansas, 1082 Malott, 1251 Wescoe Hall Dr., Lawrence, KS 66045, USA}
	
	\author{Shawn D. Domagal-Goldman}
	\affiliation{NASA Goddard Space Flight Center, 8800 Greenbelt Rd, Greenbelt, MD 20771, USA}
	
	\author[0000-0002-5967-9631]{Thomas J. Fauchez}
	\affiliation{NASA Goddard Space Flight Center, 8800 Greenbelt Rd, Greenbelt, MD 20771, USA}
	\affiliation{Goddard Earth Sciences Technology and Research (GESTAR), Universities Space Research Association (USRA), Columbia, MD 7178, USA}
	\affiliation{NASA GSFC Sellers Exoplanet Environments Collaboration}
	
	\author[0000-0001-8371-8525]{Stefano Fiscale}
	\affiliation{Dipartimento di Scienze e Tecnologia, Universit\`{a} degli Studi di Napoli Parthenope, Naples, Italy}
	
	\author{Francesco Gallo}
	\affiliation{Dipartimento di Fisica ``E. Pancini", Università di Napoli Federico II, Via Cinthia I-80126 Naples, Italy}
	
	\author[0000-0002-0388-8004]{Emily Gilbert}
	\affiliation{Department of Astronomy and Astrophysics, University of Chicago, 5640 S. Ellis Ave, Chicago, IL 60637, USA}
	\affiliation{University of Maryland, Baltimore County, 1000 Hilltop Cir, Baltimore, MD 21250, USA}
	\affiliation{The Adler Planetarium, 1300 South Lakeshore Drive, Chicago, IL 60605, USA}
	\affiliation{NASA Goddard Space Flight Center, 8800 Greenbelt Rd, Greenbelt, MD 20771, USA}
	\affiliation{NASA GSFC Sellers Exoplanet Environments Collaboration}
	
	\author[0000-0002-3385-8391]{Christina L. Hedges}
	\affiliation{Bay Area Environmental Research Institute, P.O. Box 25, Moffett Field, CA 94035, USA}
	\affiliation{NASA Ames Research Center, Moffett Field, CA 94035, USA}
	
	
	\author[0000-0002-1426-1186]{Edwin S. Kite}
	\affiliation{Department of the Geophysical Sciences, University of Chicago, Chicago, IL}
	
	\author[0000-0002-5893-2471]{Ravi K. Kopparapu}
	\affiliation{NASA Goddard Space Flight Center, 8800 Greenbelt Rd, Greenbelt, MD 20771, USA}
	
	\author[0000-0001-9786-1031]{Veselin B. Kostov}
	\affiliation{NASA Goddard Space Flight Center, 8800 Greenbelt Rd, Greenbelt, MD 20771, USA}
	\affiliation{SETI Institute, 189 Bernardo Ave, Suite 200, Mountain View, CA 94043, USA}
	\affiliation{NASA GSFC Sellers Exoplanet Environments Collaboration}
	
	
	
	\author[0000-0002-4404-0456]{Caroline Morley}
	\affiliation{Department of Astronomy, University of Texas at Austin, Austin, TX, USA}
	
	\author[0000-0001-7106-4683]{Susan E. Mullally}
	\affiliation{Space Telescope Science Institute, 3700 San Martin Dr., Baltimore, MD 21218, USA}
	
	\author[0000-0001-9771-7953]{Daria Pidhorodetska}
	\affiliation{Department of Earth and Planetary Sciences, University of California, Riverside, CA, USA}
	
	\author[0000-0001-5347-7062]{Joshua E. Schlieder}
	\affiliation{NASA Goddard Space Flight Center, 8800 Greenbelt Rd, Greenbelt, MD 20771, USA}
	
	
	\author[0000-0003-1309-2904]{Elisa V. Quintana}
	\affiliation{NASA Goddard Space Flight Center, 8800 Greenbelt Rd, Greenbelt, MD 20771, USA}
	
	\begin{abstract}
	With the increasing number of planets discovered by TESS, the atmospheric characterization of small exoplanets is accelerating. L98-59 is a M-dwarf hosting a multi-planet system, and so far, four small planets have been confirmed. The innermost planet b is $\sim15\%$ smaller and $\sim60\%$ lighter than Earth, and should thus have a predominantly rocky composition. The Hubble Space Telescope observed five primary transits of L98-59~b in $1.1-1.7\ \mu$m, and here we report the data analysis and the resulting transmission spectrum of the planet. We measure the transit depths for each of the five transits and, by combination, we obtain a transmission spectrum with an overall precision of $\sim20$ ppm in for each of the 18 spectrophotometric channels. With this level of precision, the transmission spectrum does not show significant modulation, and is thus consistent with a planet without any atmosphere or a planet having an atmosphere and high-altitude clouds or haze. The scenarios involving an aerosol-free, H$_2$-dominated atmosphere with H$_2$O or CH$_4$ are inconsistent with the data. The transmission spectrum also disfavors, but does not rules out, an H$_2$O-dominated atmosphere without clouds. \edit1{A spectral retrieval process} suggests that an H$_2$-dominated atmosphere with HCN and clouds or haze may be the preferred solution, but this indication is non-conclusive. Future James Webb Space Telescope observations may find out the nature of the planet among the remaining viable scenarios.
	\end{abstract}
	
	
	\section{Introduction} \label{sec:intro}
	
	Observational studies of the atmospheres on small and predominantly rocky exoplanets are picking up speed. The Hubble Space Telescope (HST) has measured the transmission spectra of several approximately Earth-sized exoplanets \citep[e.g.,][]{de2018atmospheric,zhang2018near,swain2021detection,mugnai2021ares} and these measurements provide important context for planning observations with other telescopes such as James Webb Space Telescope (JWST). On larger and volatile-rich exoplanets, HST was able to detect multiple chemical compounds such as H$_2$O, CH$_4$, TiO, and VO \citep[e.g.,][]{Knutson2007, Knutson2014, Swain2008, Swain2009, Fraine2014, Kreidberg2014, Evans2016, Sing2016, Tsiaras2016B2016ApJ...820...99T, Tsiaras2018, Damiano2017, Tsiaras2019,Benneke2019b}. Together with the continuing planet discoveries by the Transit Survey Exoplanet Satellite (TESS, \cite{Ricker2015}), we expect transit observations with HST and JWST to push the frontier of exoplanet atmospheric characterization to Earth-sized and likely rocky planets.
	
	L98-59 is an M3V star hosting a planetary system of four planets \citep{Cloutier2019, Damangeon2021}. Hints of a possible fifth planet have been observed but it has not yet been confirmed.
	The innermost planet, L98-59~b, is a rocky planet 15\% smaller in size than the Earth. The planet completes its orbit around its star in $\sim$2.25 days and its equilibrium temperature is $\sim627$ K as it receives more irradiation compared to Earth (see Tab.~\ref{tab:param} for the system parameters adopted in this work). High-precision radial-velocity measurements have determined that the planet has a mass of $\sim40\%$ Earth's, and the planet's mass and radius is consistent with interior-structure scenarios that range from a rocky body with no atmosphere to a planet with small but substantial gas layers \citep{Damangeon2021}. We are thus motivated to find out whether the planet has an atmosphere through spectroscopic observations.
	    
	    \begin{table*}
		\small
		\center
		\caption{Parameters of L98-59 and the planet b. The equilibrium temperature, $T_\mathrm{eq}$, is \edit1{calculated by assuming} zero albedo and efficient heat redistribution. \edit1{The parameters used to calculate $T_\mathrm{eq}$ have been adopted from \cite{Damangeon2021}.}}
		\label{tab:param}
		\begin{tabular}{c | c c}
			
			\hline \hline
			\multicolumn{3}{c}{Stellar parameters (L98-59)} \\ [0.1ex]
			\hline
			$T_\mathrm{eff}$\,[K]		& 3412 $\pm$ 49	& \edit1{\citep{Cloutier2019}} \\
			$M_* \, [M_{\odot}]$		& 0.312 $\pm$ 0.031	& \edit1{\citep{Cloutier2019}} \\
			$R_* \, [R_{\odot}]$		& 0.314 $\pm$ 0.014	& \edit1{\citep{Cloutier2019}} \\
			$[Fe/H]$ \,[dex]				& -0.5 $\pm$ 0.5 & \edit1{\citep{Cloutier2019}} \\
			$\log(g_*)$\,[cgs] 		& 4.94 $\pm$ 0.06  & \edit1{\citep{Cloutier2019}} \\ [1.0ex]
			
			\hline \hline
			\multicolumn{3}{c}{Planetary parameters (L98-59~b)} \\ [0.1ex]
			\hline
			$T_\mathrm{eq}$\,[K] 	& 627 $\pm$ 35 & \\
			$M_\mathrm{p} \, [M_\oplus]$	& 0.40 $\pm$ 0.16 & \edit1{\citep{Damangeon2021}}\\
			$R_\mathrm{p} \, [R_\oplus]$	& 0.85$\pm$ 0.054 & \edit1{\citep{Damangeon2021}}\\
			$a$\,[AU] 			& 0.02191 $\pm$ 0.00082 & \edit1{\citep{Damangeon2021}}\\ [1.0ex]
			
			\hline\hline
			\multicolumn{3}{c}{Transit parameters (L98-59~b)} \\ [0.1ex]
			\hline
			$T_0$\,[JD]			& 2458366.17067 $\pm$ 0.00035 & \edit1{\citep{Damangeon2021}}\\
			Period\,[days] 		& 2.2531136 $\pm$ 0.0000014 & \edit1{\citep{Damangeon2021}}\\
			$R_\mathrm{p}/R_*$ 	& 0.02512 $\pm$ 0.00068 & \edit1{\citep{Damangeon2021}}\\
			$i$\,[deg] 			& 87.7 $\pm$ 0.8 & \edit1{\citep{Damangeon2021}}\\ [1.0ex]
			\hline\hline
		\end{tabular}
	\end{table*}
	    
	HST has observed five primary transits of L98-59~b in the near-IR with the Wide Field Camera 3 (WFC3). In this letter, we report the data analysis and the extraction of the 1D transmission spectrum, and explore the possible atmospheric scenarios by interpreting the spectrum with a spectral retrieval algorithm. The letter is organized as follows: in Section~\ref{sec:model}, we describe two independent analyses that were used to reduce the data and extract the transmission spectrum. In Section~\ref{sec:result}, we report the extracted 1D spectra and the transit depths obtained from fitting the transit spectral and white light-curves. In Sec.\ref{sec:discussion}, with a spectral analysis, we discuss the range of the atmospheric scenarios allowed by the spectrum and how future observations may further characterize the planet. We conclude in Section~\ref{sec:conclusion}.
	    
	\section{Methods} \label{sec:model}
    \subsection{Observations}
    
    Five primary transits of L98-59~b have been observed by HST from February 2020 to February 2021 and the data are available from the MAST archive (Program ID: 15856, PI: T. Barclay, \edit1{Dataset: \dataset[10.17909/cphn-nc88]{http://dx.doi.org/10.17909/cphn-nc88}}).
    HST recorded the spatially scanned spectroscopic images through the G141 grism. Each of the five visits contained four consecutive HST orbits and each exposure was recorded in a 522$\times$522 pixels image with an exposure time of 69.617 sec each. With this configuration, the maximum signal level was 2.6$\times$10$^4$ electrons per pixel and the total scan length was approximately 300 pixels.
    The dataset also contained, for calibration purposes, a non-dispersed (direct) image of the target, obtained using the F130N filter.
    
    \subsection{Data analysis}
    
    We have used two independent data reduction and analysis pipelines to analyze the dataset.
    
    \textbf{Analysis A.} We first used \iraclis, which has been widely applied to analyze the transit observations of HST/WFC3 \citep[e.g.,][]{Tsiaras2016B2016ApJ...832..202T, Tsiaras2016B2016ApJ...820...99T, Tsiaras2018, Damiano2017}. The data reduction starts from the raw images and corrects for the bias, dark current, flat field, gain, sky background, and bad pixels. The images are then calibrated with a wavelength mapping and the signal is extracted from each image through a 2D fitting. As commonly done with WFC3 data, the images of the first HST orbit of each visit are discarded to minimize detector systematics \citep[e.g.][]{Deming2013, Huitson2013, Haynes2015, Damiano2017}. We have also discarded the first scan of each of the remaining orbits. The sequence of the signals extracted from the images composes the white and spectral (when a wavelength binning is taken into account) light-curves.
    
    We employed a parametric fitting for the white light-curve. In particular, the instrumental systematics (known as ``ramps'', \cite{Kreidberg2014,Tsiaras2016B2016ApJ...820...99T, Tsiaras2016B2016ApJ...832..202T, Damiano2017}) that affect the WFC3 infrared detector and the light-curve model are fitted at the same time to the observed data to correct for systematics and calculate the transit depths. For the correction of the ramps we used an approach similar to \citet{Kreidberg2014}, i.e., adopting an analytic function with two different types of ramps, short-term and long-term:
        \begin{equation}
	    R(t) = (1-r_{a}(t-t_{v}))(1-r_{b1}e^{-r_{b2}(t-t_{0})})
	    \label{rampfunction}
	    \end{equation}
	where $t$ is the mid-time of each exposure, $t_v$ is the time when the visit starts, $t_0$ is the time when each orbit starts, $r_a$ is related to the long-term ramp, and $r_{b1}$ and $r_{b2}$ are related to the short-term ramp. This systematics function together with the light-curve model and the limb darkening coefficients previously calculated are used as the fitting model for the observed white light-curves: 
	  
	  \begin{equation}
	    M(t) = n_w R(t) F(t)
	  \end{equation}
	
	\noindent where $t$ represents the time, $R(t)$ is the systematics function (Eq. \ref{rampfunction}), $F(t)$ is the transit model calculated by using \texttt{pylightcurve} \citep{Tsiaras2016B2016ApJ...832..202T}, and $n_w$ is a normalization factor. \edit1{Moreover, we note that due to the HST spatial scanning techniques (i.e. the telescope slews slowly along the cross-dispersion direction instead of staring at the target), the images related to the downward scanning direction have a different associated normalization factor compared to those recorded with an upward scanning \citep{Tsiaras2016B2016ApJ...832..202T}. Therefore, the normalization factors are $n_{w,for}$ and $n_{w,rev}$ for forward and reverse scanning respectively}.
	
	\edit1{To account for the limb darkening effect, we adopted the Claret's formulation \citep{Claret2000} (4-parametric expression). The coefficients have been calculated by using \exot\ \citep{Morello2020} and provided in Appendix~\ref{sec:limb}.} 
	
	We used Markov Chain Monte Carlo (MCMC) statistical tool \edit1{(the \texttt{python} package \texttt{emcee} \citep{Foreman-Mackey2013})} to fit the five white light-curves. 200 walkers have been deployed and 500,000 iterations have been considered (the first 200,000 iterations are considered to be burned to allow the chains to stabilize). \edit1{The free parameters of the fitting are: the two normalization factors $n_{w,for}$ and $n_{w,rev}$, the three parameters describing the instrumental systematics in Eq.~\ref{rampfunction}, $r_a$, $r_{b1}$, and $r_{b2}$, the planetary and stellar radius ratio, $R_p/R_{\star}$, and finally, the mid-transit time, $T_0$.}
	
	Lastly, to fit the spectral light-curves, for each wavelength bin we divided the spectral light-curve by the white light-curve \citep{Kreidberg2014} and fitted a linear trend simultaneously with a relative transit model. Also in this case we used MCMC to fit the model to the data points. For the spectral light-curve, we used 100 walkers and considered only 50,000 iteration (first 20,000 were considered burned iterations) as the complexity of the fitting is lower than the white light-curve. \edit1{When the spectral light-curves are divided by the white one, the effect of the instrumental systematics, described in Eq.~\ref{rampfunction}, is also divided out. As such, the reduction of the number of iterations does not impact the convergence of the parameters. For the spectral light-curves fitting the free parameters are: the two normalization factors of relative to forward and reverse scanning, $n_{w,for}$ and $n_{w,rev}$, the planetary and stellar radius ratio, $R_p/R_{\star}$, and finally, the mid-transit time, $T_0$.}
	
	    
	\textbf{Analysis B.} We also used a custom pipeline described in \citet{Kreidberg2014} and \citet{Kreidberg2018} to reduce the \texttt{ima} data products which we accessed from the MAST archive. The \texttt{ima} (intermediate MultiAccum) files already had all calibrations applied (dark subtraction, and linearity correction, flat-fielding) to each readout of the IR exposure. Every orbit of the observations started with an direct image, which we used to determine centroid position of the star on the detector. 

    We separately extracted each up-the-ramp sample and subtracted the background flux from them. The background flux was determined by taking the median flux of the pixels where the spectrum did not fall on. We used the optimal extraction routine presented in \citet{Horne1986} to extract the spectra and then coadded the individual samples to get the final spectrum for each exposure. To correct for spectral drift, we cross-correlated the first exposure in every orbit with a reference spectrum consisting of the product of the bandpass of the WFC3/G141 instrument and a \texttt{PHOENIX} \edit1{\footnote{\url{https://archive.stsci.edu/hlsps/reference-atlases/cdbs/grid/phoenix/}}} stellar model \citep{Allard2003} for L98-59~b. \edit1{We used the PHOENIX stellar model which was the closest to the stellar parameters from \citep{Cloutier2019}, i.e., the one corresponding to $T_{eff}$ $=$ 3400K, log(g) $=$ 5.00 and MH $=$ -0.5.} 
    Due to strong ramp-like features in the raw data caused by charge traps filling up in the detector \citep{Zhou2017}, we also removed the first orbit in every visit and the first exposure in every orbit, as in previous WFC3 analysis \citep{Kreidberg2014}.

    Our fitting model $F_{B}(t)$ consists of a transit model $F_{\rm{transit}} (t)$ which is implemented in the open-source python package \texttt{batman} \citep{Kreidberg2015} and a systematic model $F_{\rm{sys}} (t)$ to fit for the WFC3 systematics:
    \begin{equation}
        F_B(t)=F_{\rm{transit}} (t) \, F_{\rm{sys}} (t).
    \end{equation}

    The systematic model $F_{\rm{sys}} (t)$ consists of a visit long linear trend $F_{\rm{sys, visit}} (t)$ and an exponential ramp for each orbit $F_{\rm{sys, orbit}} (t)$:
    \begin{equation}
    \begin{split}
     F_{\rm{sys}}(t) &= F_{\rm{sys, visit}} (t) \, F_{\rm{sys, orbit}} (t)\\
     &= (c\,S(t) + k\,t_{\rm{v}}) \, (1 - \exp(-r_1\,t_{\rm{orb}} - r_2 - r_3\,t_{\rm{orb}}\,Q(t))),
    \end{split}
    \end{equation}
    where $t_v$ is the time since the first exposure in a visit and $t_{orb}$ is the time since the first exposure in an orbit. The linear trend $F_{sys, visit} (t)$ includes the flux constant $c$ and the slope $k$. $S(t)$ accounts for the upstream-downstream effect \citep{McCullough2012} which leads to an alternating total flux between exposures with spatial scanning in the forward direction and exposures with reverse scans. We define this scale factor to be $S(t) = 1$ for forward scans and $S(t) = s$ for reverse scans. The exponential ramp parameters are $r_1$, $r_2$ and $r_3$. Because the first remaining orbit (hereafter ``first orbit") in every visit exhibited a stronger exponential ramp than the following ones, we included a rectangular function $Q(t)$, where $Q(t) = 1$ for the first orbit in a visit, and $Q(t) = 0$ for the others. For the fits we allowed all systematic parameters ($c$, $k$, $s$, $r_1$, $r_2$ and $r_3$) to have different values from visit to visit. For the spectroscopic light curve fits, we additionally allowed these parameters to vary for every spectroscopic bin.
    
    
    \edit1{For the white light-curves fit and the spectroscopic fits we fixed the orbital period $P$, ratio of semi-major axis to stellar radius $a/R_s$ and the orbital inclination $i$ to literature values \citep{Damangeon2021}. We also fixed the eccentricity to zero. For the white light-curves fit we therefore fitted for the ratio of planet to stellar radius $R_p/R_s$, the transit time $T_0$ for each visit, a linear limb darkening parameter $u_1$ and for the six different systematic parameters ($c$, $k$, $s$, $r_1$, $r_2$ and $r_3$) for each visit. For the spectroscopic light-curve fits, we fixed the transit time $T_0$ of each visit to the best-fit values from the white light-curve fit.}
    
    We used the MCMC Ensemble sampler package \texttt{emcee} \citep{Foreman-Mackey2013} to estimate the parameters and their uncertainties for our model. We rescaled the uncertainties for every data point by a constant factor so that the reduced chi-squared is unity, to ensure we are not underestimating the uncertainties of our parameters. For the white light curve and each spectroscopic light curve, we ran 12000 steps and 80 walkers and disregarded the first half of the samples as burn-in. \edit1{Finally, in Analysis B, we decided not to fit each visit's white light-curve independently; instead, we fitted them within the same instance by fitting for one $R_p/R_s$ and five mid-transit time $T_0$. Similarly, the spectral light-curves for each wavelength bin from all visits have been fitted together.}
    
	\section{Results} \label{sec:result}
	
\begin{figure*}[!h]
		\centering
		\includegraphics[angle=0, scale=0.90]{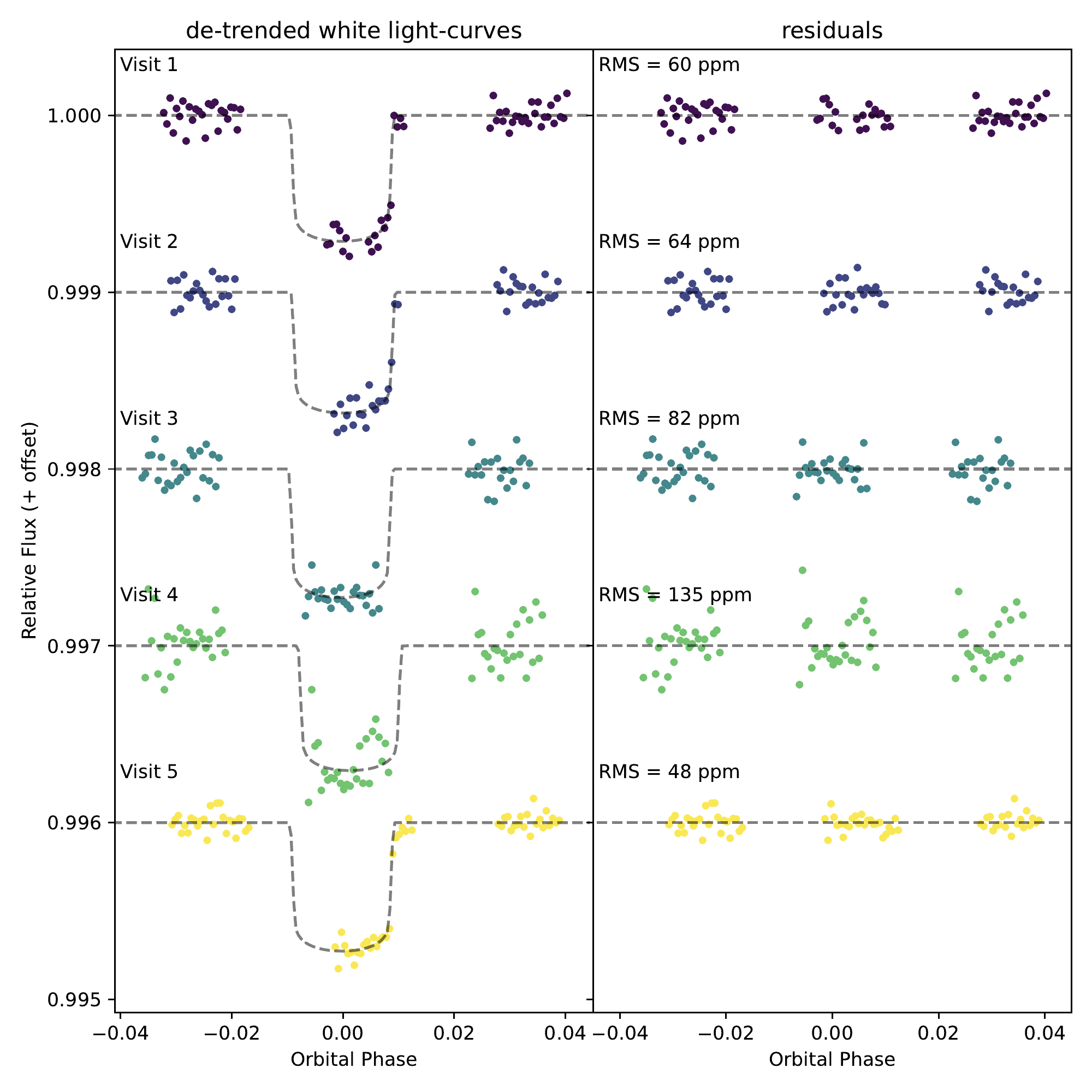}
		\caption{\edit1{White light-curves of HST observations of L98-59~b transits. For each visit, the left panel shows the light-curves from 2D images after correcting for systematics, and the right panel shows the residuals of the fit of the white light-curves. The dashed line depicts the best-fit transit model. A vertical shift of $0.001$ is applied between the visits to better visualize the curves. \label{fig:lw_single}}}
	\end{figure*}

 \edit1{Fig.~\ref{fig:lw_single} presents white light-curves corrected for instrument systematics and fitted by the transit model, based on Analysis A. The mid-transit time and R$_p$/R$_s$ are reported in Tab.~\ref{tab:wlight_fit}. The residuals do not show correlated signals. We observed an increased scatter in the light-curve of Visit 4, which resulted in greater uncertainty in R$_p$/R$_s$.}	
 

\edit1{The derived 1D spectra from Analysis A are reported in Tab.~\ref{tab:resultsA}, and the quality of the spectral light-curve fits is reported in Appendix~\ref{sec:single_visit}. We observe that the standard deviation of the residuals is close to the photon-noise limit, supporting the practice to divide the spectral light-curves by the white ones to remove instrumental systematics.}
 

\edit1{From Analysis B, we obtained a white light-curve simultaneous fit to the five visits with an RMS of 48 ppm. The mid-transit time and R$_p$/R$_s$ are also reported in Tab.~\ref{tab:wlight_fit}, and they are comparable to the result from Analysis A. The derived 1D spectrum is reported in Tab.~\ref{tab:resultsB}.}

\begin{figure*}[!h]
    		\centering
    		\includegraphics[width=\textwidth]{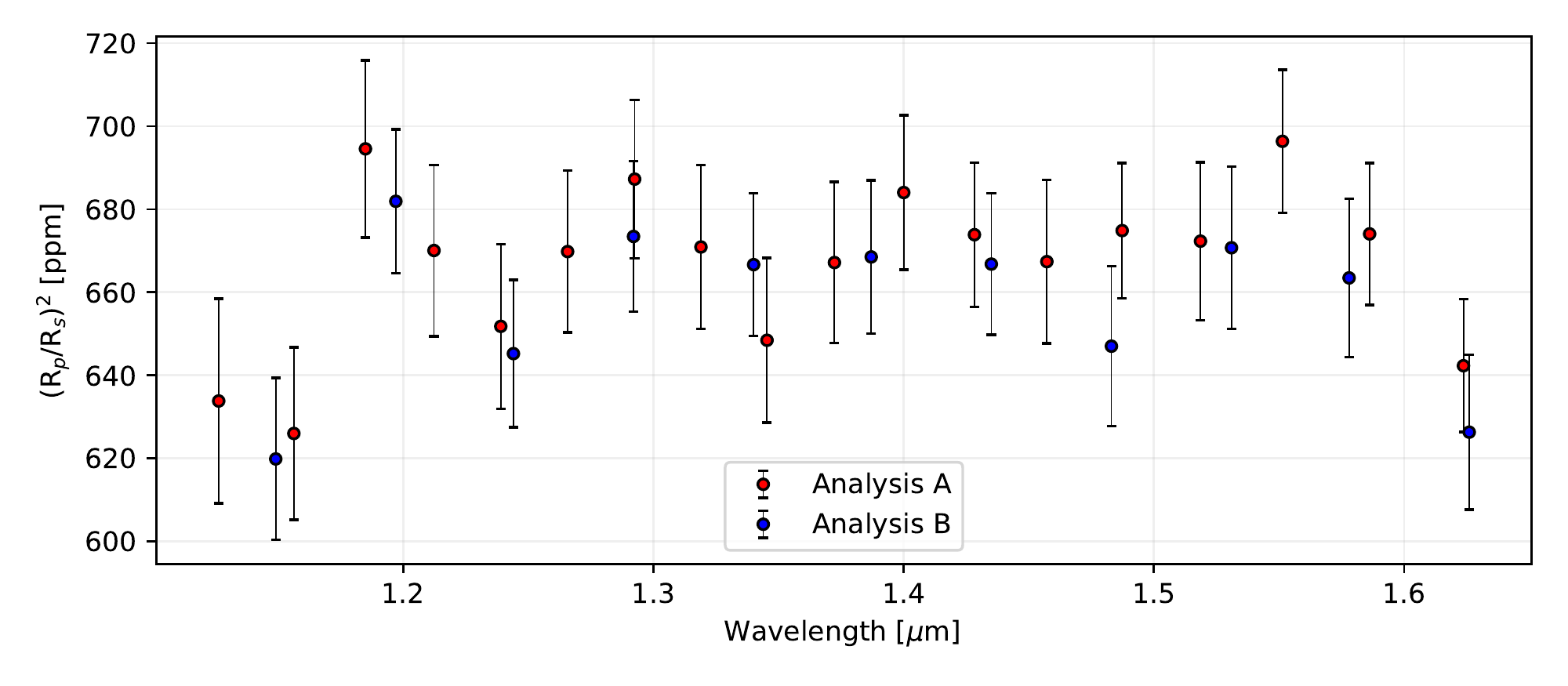}
    		\caption{Transmission spectra of L98-59~b using the two independent methods described in Section~\ref{sec:model}. Each of the two spectra is the weighted average of the five single visits. \label{fig:1d_final}}
    	\end{figure*}

\edit1{The resulting spectra from the two analyses are shown in Fig.~\ref{fig:1d_final}. For Analysis A, the weighted average of the single-visit spectra is calculated. The two methods provided consistent results within 1$\sigma$ across the wavelength range, and the uncertainties provided by the two methods are also consistent.}
 

	\begin{table*}
		\small
		\center
		\caption{\edit1{Mid-transit time and R$_p$/R$_s$ fitted from white light-curves for both analyses. For Analysis B a single value of R$_p$/R$_s$ has been fitted simultaneously for all visits, while for Analysis A, the combined value of R$_p$/R$_s$ refers to the weighted average of the single-visit values.}}
		\label{tab:wlight_fit}
		\begin{tabular}{c | c c}
		
		    \hline \hline
			Mid-point [BJD+2450000] & Analisys A & Analisys B \\
			\hline
			Visit 1 & $8888.89348\pm0.00016$ & $8888.8921\pm0.0012$ \\
			Visit 2 & $8947.47386\pm0.00029$ & $8947.4730\pm0.0012$ \\
			Visit 3 & $9120.9646\pm0.0023$ & $9120.9626\pm0.0018$ \\
			Visit 4 & $9179.542\pm0.004$ & $9179.5435\pm0.0021$ \\
			Visit 5 & $9269.66962\pm0.00010$ & $9269.6680\pm0.0012$ \\
			
			\hline \hline
            (R$_p$/R$_s$)$^2$ [ppm] & & \\
			\hline
			Visit 1 & 663 $\pm$ 21 & - \\
			Visit 2 & 635 $\pm$ 26 & - \\
			Visit 3 & 676 $\pm$ 27 & - \\
			Visit 4 & 656 $\pm$ 36 & - \\
			Visit 5 & 676 $\pm$ 16 & - \\
			\hline
			Combined & 665.2 $\pm$ 11.7 & 642.6 $\pm$ 10.8 \\
			\hline\hline
		\end{tabular}
	\end{table*}
	    
    	\begin{table*}
		\small
		\center
		\caption{Derived 1D spectra ((R$_p$/R$_s$)$^2$) from the five visits of L98-59b using the Analysis A.}
		\label{tab:resultsA}
		\begin{tabular}{c c c c c c | c}
			
			\hline \hline
			Spectral bins [nm] & V1 & V2 & V3 & V4 & V5 & Weighted Average [ppm] \\
			\hline
			1110.8 - 1141.6 & $636\pm65$ & $600\pm62$ & $670\pm57$ & $617\pm47$ & $647\pm51$ & $634\pm25$ \\
			1141.6 - 1170.9 & $626\pm63$ & $664\pm47$ & $662\pm39$ & $596\pm41$ & $567\pm52$ & $626\pm21$ \\
			1170.9 - 1198.8 & $680\pm46$ & $724\pm47$ & $676\pm48$ & $657\pm74$ & $708\pm39$ & $695\pm21$ \\
            1198.8 - 1225.7 & $636\pm38$ & $642\pm52$ & $732\pm75$ & $711\pm40$ & $666\pm45$ & $670\pm21$ \\
            1225.7 - 1252.2 & $652\pm57$ & $600\pm42$ & $631\pm38$ & $686\pm43$ & $708\pm47$ & $652\pm20$ \\
            1252.2 - 1279.1 & $670\pm44$ & $586\pm42$ & $644\pm44$ & $745\pm36$ & $673\pm62$ & $670\pm19$ \\
            1279.1 - 1305.8 & $695\pm41$ & $649\pm63$ & $626\pm46$ & $700\pm37$ & $722\pm37$ & $687\pm19$ \\
            1305.8 - 1332.1 & $679\pm40$ & $569\pm46$ & $675\pm43$ & $677\pm47$ & $749\pm45$ & $671\pm20$ \\
            1332.1 - 1358.6 & $688\pm53$ & $635\pm39$ & $671\pm49$ & $580\pm43$ & $690\pm42$ & $648\pm20$ \\
            1358.6 - 1386.0 & $667\pm57$ & $558\pm39$ & $717\pm36$ & $726\pm49$ & $686\pm45$ & $667\pm19$ \\
            1386.0 - 1414.0 & $696\pm44$ & $612\pm56$ & $710\pm35$ & $626\pm48$ & $711\pm35$ & $684\pm19$ \\
            1414.0 - 1442.5 & $666\pm60$ & $666\pm41$ & $691\pm32$ & $664\pm37$ & $670\pm36$ & $674\pm17$ \\
            1442.5 - 1471.9 & $672\pm52$ & $693\pm49$ & $661\pm37$ & $643\pm37$ & $691\pm52$ & $667\pm20$ \\
            1471.9 - 1502.7 & $659\pm40$ & $652\pm37$ & $717\pm33$ & $665\pm31$ & $670\pm48$ & $675\pm16$ \\
            1502.7 - 1534.5 & $677\pm56$ & $675\pm44$ & $722\pm42$ & $611\pm35$ & $707\pm42$ & $672\pm19$ \\
            1534.5 - 1568.2 & $683\pm38$ & $751\pm48$ & $687\pm34$ & $700\pm32$ & $675\pm48$ & $696\pm17$ \\
            1568.2 - 1604.2 & $658\pm43$ & $638\pm38$ & $680\pm33$ & $727\pm39$ & $663\pm39$ & $674\pm17$ \\
            1604.2 - 1643.2 & $665\pm49$ & $632\pm30$ & $647\pm37$ & $636\pm29$ & $654\pm45$ & $642\pm16$ \\
			\hline\hline

		\end{tabular}
	\end{table*}
	
	\begin{table}
		\small
		\center
		\caption{\edit1{Derived 1D spectrum ((R$_p$/R$_s$)$^2$) from the five visits of L98-59b using the Analysis B. The spectral light-curves of the five visits are fit together to produce the combined constraints on (R$_p$/R$_s$)$^2$ for each wavelength bin.}}
		\label{tab:resultsB}
		\begin{tabular}{c c}
			
			\hline \hline
			Spectral bins [nm] & (R$_p$/R$_s$)$^2$ [ppm] \\
			\hline
			1125.0 - 1173.0 & $620\pm20$ \\
			1173.0 - 1220.5 & $682\pm17$ \\
			1220.5 - 1268.0 & $645\pm18$ \\
            1268.0 - 1316.0 & $673\pm18$ \\
            1316.0 - 1363.5 & $667\pm17$ \\
            1363.5 - 1411.0 & $669\pm18$ \\
            1411.0 - 1459.0 & $667\pm17$ \\
            1459.0 - 1507.0 & $647\pm19$ \\
            1507.0 - 1554.5 & $671\pm20$ \\
            1554.5 - 1602.0 & $663\pm19$ \\
            1602.0 - 1650.0 & $626\pm19$ \\
			\hline\hline

		\end{tabular}
	\end{table}
	   
	\section{Discussion} \label{sec:discussion}
	    
	\subsection{Plausible planetary scenarios}
 
\begin{figure*}[!h]
    		\centering
    	\includegraphics[width=\textwidth]{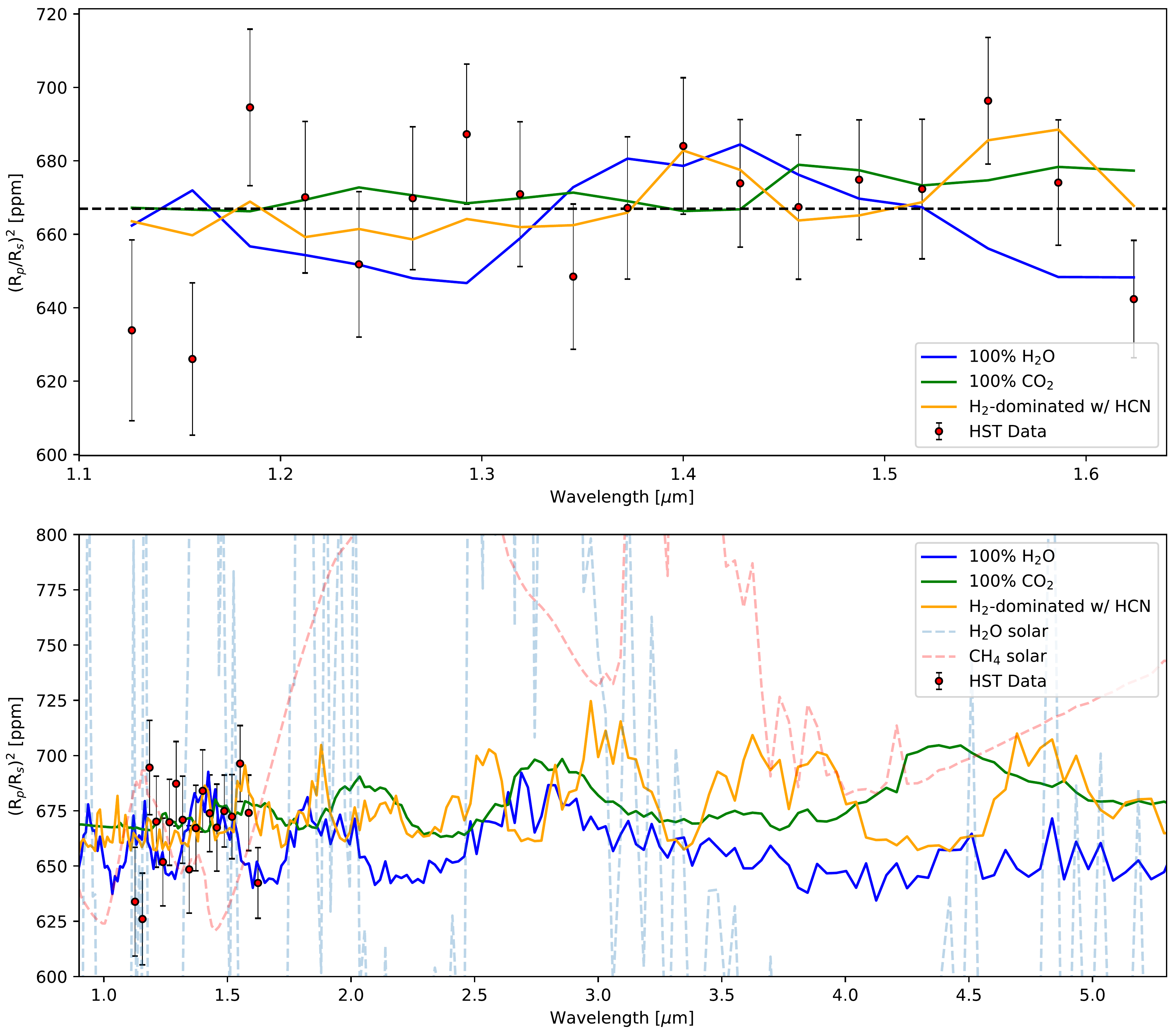}
    		\caption{\textbf{Top panel:} Transmission spectrum of L98-59~b in comparison with model scenarios. The scenarios of an H$_2$O- or CO$_2$-dominated atmosphere assume that the atmosphere does not have clouds or haze and has a temperature of the planet's equilibrium temperature. The H$_2$-dominated atmosphere model is the model calculated from the median of the posterior distribution shown in Fig.~\ref{fig:HCN_post}. This model has HCN with a volume mixing ratio of 360 parts per million and clouds at $10^3$ Pa in the atmosphere. \textbf{Bottom panel:} The HST data \citep{Tsiaras2019,Benneke2019b} comparing with atmospheric models of L98-59~b in the context of JWST observations. The dashed lines show the models with an H$_2$-dominated atmosphere and solar-abundance water and methane for comparison -- these models are ruled out by the current data. The models permitted by the transmission spectrum reported here can be distinguished with JWST observations in $\sim1.5-5\ \mu$m.} \label{fig:best_fit}
    	\end{figure*}

	The transmission spectra do not show apparent modulation. The data are consistent with a flat spectrum where the transit depth does not change with wavelength with $\chi^2=17.8$ for a degree of freedom of 17. This means that the transmission spectra are consistent with a planet without any atmosphere (i.e., a bare-rock planet), or a planet with an atmosphere and high-altitude clouds or haze. For comparison, we have calculated the spectra of cloud-free H$_2$-dominated atmospheres with solar-abundance H$_2$O or CH$_4$, and they are clearly ruled out by the observed spectra (Fig.~\ref{fig:best_fit}). \edit1{We used \taurex\ \citep{Al-Refaie2021} and the opacities included in \exotransmit \footnote{\url{https://github.com/elizakempton/Exo_Transmit/tree/master/Opac}} \citep{Freedman2008,Freedman2014,Lupu2014,Kempton2017} to synthesize the model transmission spectra.}  

We have also tested whether the transmission spectrum of L98-59b would be consistent with a high mean molecular weight atmosphere. We found the spectra to be consistent with a \edit1{cloud-free} CO$_2$-dominated atmosphere with $\chi^2=21.0$ (Fig.~\ref{fig:best_fit}). Interestingly, the transmission spectrum \edit1{does not favor an} H$_2$O-dominated atmosphere. We found that an H$_2$O-dominated atmosphere without clouds or haze would be \edit1{excluded with a p-value of 0.1 ($\chi^2=23.6$)}. The reason for this potential inconsistency is that an H$_2$O-dominated atmosphere should cause a rise in the transit depth at $\sim1.4\ \mu$m, which is absent from the data. This finding is somewhat surprising because an H$_2$O-dominated atmosphere was one of the most likely scenarios for this planet prior to the observations \citep{Damangeon2021}. On the other hand, the observed spectra can be consistent with an H$_2$O-dominated atmosphere if clouds are included \edit1{($\chi^2=18.1$)}, similar to the case of GJ~1214~b \citep{Kreidberg2014}.
	
	To explore the potential atmospheric scenarios of the planet in a more systematic way, we used \taurex\ \citep{Al-Refaie2021} to run a statistical inverse process to reveal the range of atmospheric conditions that would be consistent with the observed transmission spectrum. We adopted the results from Analysis A in the spectral retrieval. We assumed that the background atmosphere was dominated by molecular hydrogen and helium, but also allowed any gas of interest to take a mixing ratio of (almost) unity, effectively allowing, for example, an H$_2$O-, CO$_2$- or N$_2$-dominated atmosphere. We considered a broad range of molecules as candidate trace gases, including H$_2$O, CH$_4$, CO$_2$, CO, and HCN. Given the relatively narrow spectral range probed, we assumed an isothermal temperature profile and molecular abundances constant with pressure. We set \edit1{log- or linear-}uniform priors to the fitted parameters, which are: the \edit1{log} mixing ratios of the molecules ($log(10^{-12}) - log(10^{-0.1})$), the temperature (100 -- 800 K), the radius of the planet (0.038 -- 0.15 $R_\mathrm{Jup}$), and the cloud top pressure ($10^{-3} - 10^7$ Pa). The retrieval model additionally has the opacity contribution from Rayleigh scattering and collision-induced absorption from H$_2$-H$_2$ and H$_2$-He pairs. \edit1{We used \multinest \citep{Feroz2007,Feroz2009,Feroz2013,Buchner2014} as optimizer of the retrieval, which is the implementation of the \textit{nested sampling} algorithm \citep{Skilling2004,Sivia2006,Skilling2006}. We set the number of live points to 600 which is safely higher than four times the number of free parameters which are nine in total. Finally, we set the evidence tolerance to the standard value 0.5.} 
 
 The spectral retrieval suggested an interesting scenario that involves an H$_2$-dominated atmosphere, clouds at $\sim10^3$ Pa, and substantial presence of HCN (see Appendix \ref{sec:HCN} for the posterior distributions). The volume mixing ratio (VMR) of HCN is not well-constrained, but  rather, the likelihood for higher values (\textgreater 10$^{-2.5}$) is greater than lower values. \edit1{The model spectrum referring the median of posterior distributions that includes an H$_2$-dominated atmosphere with HCN and clouds is shown in Fig. \ref{fig:best_fit}. For this scenario, we report a Bayesian log-evidence of 165.27 $\pm$ 0.09.}
Meanwhile, the retrieval disfavors any presence of H$_2$O or CH$_4$, and does not yield any constraints on CO, CO$_2$, or N$_2$ (Appendix \ref{sec:HCN}). \edit1{If HCN is not included as a candidate molecule in the retrieval, the retrieval converges to a flat spectrum with a Bayesian log-evidence of 164.87 $\pm$ 0.09, slightly lower than the case with HCN. For completeness we have also run the retrieval for all the model discussed in this section, and the Bayesian log-evidence results are reported in Tab. \ref{tab:logev}. The relative values of the Bayesian evidence are consistent with the $\chi^2$ metrics reported earlier.}

\begin{table*}
		\small
		\center
		\caption{\edit1{Bayesian log-evidence, log(EV), resulting from the retrieval process to the 1D spectrum when different atmospheric scenarios are considered.}}
		\label{tab:logev}
		\begin{tabular}{c c}
			
			\hline \hline
			Scenario & log(EV) \\
			\hline
			H$_2$-dominated atmosphere $+$ clouds $+$ HCN & 165.44 $\pm$ 0.09 \\
			H$_2$-dominated atmosphere $+$ clouds $+$ H$_2$O $+$ CH$_4$ $+$ HCN $+$ CO$_2$ $+$ CO $+$ N$_2$ & 165.27 $\pm$ 0.09 \\
			Fully clouded - flat spectrum & 164.87 $\pm$ 0.09 \\
            CO$_2$-dominated atmosphere cloud free & 164.86 $\pm$ 0.09 \\
            H$_2$O-dominated atmosphere $+$ clouds & 164.86 $\pm$ 0.09 \\
            H$_2$O-dominated atmosphere cloud free & 160.93 $\pm$ 0.09 \\
			\hline\hline

		\end{tabular}
	\end{table*}

    \subsection{Stellar activity}
    
    While the initial discovery paper for the L98-59 planets indicated that no stellar variability was detected \citep{Kostov2019}, at least one flare is seen in subsequent TESS observations, and evidence of activity was identified in radial velocity data observations \citep{Cloutier2019,Damangeon2021}. Stellar activity can potentially mimic or mask the detection of at atmospheric signal in transmission spectra \citep{Pont2008,Bean2010,Sing2011,Aigrain2012,Huitson2013,Jordan2013,Kreidberg2014,McCullough2014,Barstow2015,Nikolov2015,Herrero2016,Zellem2017,Rackham2018,Rackham2019,Barclay2021}. This is a particular challenge for early-mid M-dwarf stars where the residuals from stellar H$_2$O absorption can cause issues in the interpretation of data from $1.1-1.7\ \mu$m.
    The rotation rate of L98-59 is slow, likely in the region of 80 days \citep{Cloutier2019,Damangeon2021}. Therefore, the star rotates minimally during the 0.9 hour transit of L98-59~b. This makes a potential contamination signal somewhat less likely. However, even slowly rotating spotted stars are not immune to contaminated transmission spectra owing to the transit light source effect \citep{Rackham2018}. 

    All of the five observations of L98-59~b have consistent transmission spectra, which provides confidence that a single observation does not dominate the combined data and bias the conclusions (Appendix~\ref{sec:single_visit}). Furthermore, the characteristic bump of a contaminated spectrum at 1.4 $\mu$m due to H$_2$O in the stellar atmosphere \citep{Barclay2021} is not observed for this planet. 
    Therefore, while we cannot rule out that our observations contain some level of uncorrected stellar signal, we do not see any evidence for this.

	\subsection{Prospects of future studies}
	
	L98-59~b is currently the lowest-mass exoplanet measured through stellar radial velocities, and the uncertainties of the measured planetary mass and radius permit the planetary scenarios of a rocky body without any substantial gas or ice, a planet with $\sim20\%$ water by mass, or a planet with a small H$_2$/He gas layer \citep{Damangeon2021}. It is thus particularly interesting to find out whether the planet has an atmosphere. Basic theoretical models suggest that sub-Earths in 2-day orbits should not have an atmosphere \citep[e.g.,][]{Zahnle2017}. However, these models do not include many effects that might allow small, highly irradiated planets to exhibit an atmosphere, for example the low escape efficiency for CO$_2$-dominated atmospheres \citep{Tian2009, Johnstone2021}. It is also plausible that the planet has retained some volatile through the early evolution \citep[e.g.,][]{Kite2021} or has a secondary atmosphere from volcanic outgassing \citep[e.g.,][]{kite2020exoplanet}. However, non-thermal escape processes may be very effective in removing the secondary atmosphere \citep[e.g.,][]{dong2018atmospheric}, and moreover, volcanic outgassing may shut down quickly on sub-Earth-mass planets \citep{Kite2009}. 
	
	If the planet does not have an atmosphere, its rocky surface could be spectroscopically detectable via Si-O features in $7-13\ \mu$m \citep{hu2012theoretical}. Similar to the case of LHS~3844~b \citep{kreidberg2019absence}, thermal emission spectroscopy using JWST's MIRI instrument may detect the signatures of ultramafic, basaltic, and granitoid surfaces on this planet and reveal its geologic histories.
	
	If the planet is a water world, it should have a steam atmosphere given the level of irradiation \citep[e.g.,][]{turbet2020revised}. A cloudless H$_2$O atmosphere is not favored by the transmission spectra reported here, but is not ruled out. Theoretical calculations suggest that the escape efficiency for pure-H$_2$O atmospheres is high, similar to that for pure-H$_2$ atmospheres, disfavoring retention of a pure-H$_2$O atmosphere \citep{Johnstone2020}. A CO$_2$-dominated atmosphere built up by volcanic outgassing would be consistent with the transmission spectra. Because of L98-59~b's low mass, this scenario would require that the planet start with much more CO$_2$ per unit mass than Earth or Venus according to the modeling of \citet{kite2020exoplanet}. Therefore, if a CO$_2$-dominated atmosphere is detected in future, it would have major implications for the distribution and radial transport of Life-Essential Volatile Elements (LEVEs) including carbon to distances close to the star \citep{dasgupta2019origin}. The photometric precision achieved in this study is $\sim20$ ppm per spectral channel, and a substantial improvement over this using HST may be very hard or inefficient. Fig.~\ref{fig:best_fit} suggests that a precision of $\sim20$ ppm in $1.5-5\ \mu$m at a moderate spectral resolution, which should be within the reach of the instruments on JWST \cite[e.g.,][]{beichman2014observations}, could detect such a non-H$_2$-dominated atmosphere on the planet and characterize its bulk composition. Because of L98-59~b's small radius and known mass, such observations could provide a particularly powerful test for the models of planet-mass-dependent atmospheric retention and evolution on small planets.
	
	Lastly, let us consider an H$_2$-dominated atmosphere having HCN and clouds. This scenario is favored by the retrieval of the transmission spectra as the statistical convergence tries to find the best model that fits the bump at 1.55 $\mu$m. The retrieval selects HCN in an H$_2$-dominated atmosphere and then it also invokes a cloud deck to produce an otherwise flat spectrum. This H$_2$-dominated atmosphere cannot be massive because, for the equilibrium temperature of $\sim600$ K, an H$_2$-dominated atmosphere would already have a thickness of $0.4\ R_{\oplus}$ from 0.001 to 1 bar. To keep the H$_2$-dominated atmosphere small but existing is a fine-tuning problem as there is no known feedback mechanism that stabilizes the mass of the atmosphere. Is a small H$_2$-dominated atmosphere a plausible scenario from the atmospheric evolution point of view? It is possible that most of the initial endowment of hydrogen has been lost during early evolutions \citep[e.g,][]{misener2021cool}, and most of the remaining hydrogen is partitioned into the magma ocean \citep{kite2020atmosphere,gaillard2022redox}. Volcanoes can release H$_2$-dominated gases into the atmosphere from the mantle \citep{liggins2020can}, and outgassing from impactors may also release H$_2$-dominated gases \citep{schaefer2017redox}. Therefore, it would be worthwhile to pursue the hint of HCN suggested in this scenario. HCN is a common photochemical product in temperate, H$_2$-dominated atmospheres \citep{hu2021photochemistry} as well as in hot, N$_2$-dominated atmospheres \citep{miguel2019observability}. Warm atmospheres on rocky exoplanets with volcanic outgassing could also have HCN \citep{swain2021detection}. There have been debated reports of HCN from exoplanet transit observations \citep{Tsiaras2016B2016ApJ...820...99T,swain2021detection}, and here, the rise of the transit depth between $1.5-1.6\ \mu$m has been found by two data analyses and is most naturally explained as the HCN absorption (Fig.~\ref{fig:best_fit}). This HCN, along with other possible scenarios discussed above, could be confirmed or refuted by observing the planet at longer wavelengths. 
	
	The other two detected planets in the L98-59 system (planets c and d) have been observed by HST within the program 15856 and the findings will be reported in a separate paper (Barclay et al. in prep). Moreover, JWST will observe the planets c and d in $0.6-5\ \mu$m through multiple programs in Cycle 1. The L98-59 system is poised to become one of best characterized exoplanetary systems with multiple small planets. Comparing the transmission spectra of the three planets could reveal system-wide trends in the atmospheric composition and thus volatile retention. 
	
	
	\section{Conclusion} \label{sec:conclusion}
	
	In the paper, we report the transmission spectra of the warm sub-Earth-sized exoplanet L98-59~b in $1.1-1.7\ \mu$m, obtained by multiple-visit observations of the HST. We applied two independent data analysis pipelines and obtained consistent results. Combining five visits, we achieved a photometric precision of $\sim20$ ppm per spectral channel, with 18 channels in $1.1-1.7\ \mu$m ($R\sim50$), making the transmission spectrum reported here one of the most precise measurements from an exoplanet \citep[e.g.,][]{Kreidberg2014,Tsiaras2019,Benneke2019b}. The spectrum does not show significant modulation, and thus rules out a cloud-free H$_2$-dominated atmosphere with solar abundance of H$_2$O or CH$_4$. The spectrum also does not favor a cloud-free H$_2$O-dominated atmosphere. In addition to the null hypothesis (i.e., a bare-rock planet or an atmosphere with high-altitude clouds or haze), the spectrum is consistent with a \edit1{cloud-free} CO$_2$-dominated atmosphere or a small H$_2$-dominated atmosphere with HCN and clouds/haze. JWST observations of the planet at the precision of $\sim20$ ppm per spectral channel in a wide wavelength range could test these atmospheric scenarios and thus determine the nature of the planet. As a sub-Earth-sized planet, L98-59~b provides a valuable opportunity to test the volatile retention and evolution on small and irradiated exoplanets.
	
	\section*{Acknowledgments}
	
    We thank A. Youngblood for assistance with the HST observations. This research is based on observations made with the NASA/ESA Hubble Space Telescope obtained from the Space Telescope Science Institute, which is operated by the Association of Universities for Research in Astronomy, Inc., under NASA contract NAS 5–26555. These observations are associated with program 15856. Support for program \#15856 was provided by NASA through a grant from the Space Telescope Science Institute. This work was supported by the GSFC Sellers Exoplanet Environments Collaboration (SEEC), which is funded by the NASA Planetary Science Divisions Internal Scientist Funding Mode. The material is based on work supported by NASA under award No. 80GSFC21M0002. Part of the research was carried out at the Jet Propulsion Laboratory, California Institute of Technology, under a contract with the National Aeronautics and Space Administration.
	
	{	\small
		\bibliographystyle{apj}
		\bibliography{bib.bib}

\begin{thebibliography}{}
\expandafter\ifx\csname natexlab\endcsname\relax\def\natexlab#1{#1}\fi
\providecommand{\url}[1]{\href{#1}{#1}}
\providecommand{\dodoi}[1]{doi:~\href{http://doi.org/#1}{\nolinkurl{#1}}}
\providecommand{\doeprint}[1]{\href{http://ascl.net/#1}{\nolinkurl{http://ascl.net/#1}}}
\providecommand{\doarXiv}[1]{\href{https://arxiv.org/abs/#1}{\nolinkurl{https://arxiv.org/abs/#1}}}

\bibitem[{{Aigrain} {et~al.}(2012){Aigrain}, {Pont}, \& {Zucker}}]{Aigrain2012}
{Aigrain}, S., {Pont}, F., \& {Zucker}, S. 2012, \mnras, 419, 3147,
  \dodoi{10.1111/j.1365-2966.2011.19960.x}

\bibitem[{{Al-Refaie} {et~al.}(2021){Al-Refaie}, {Changeat}, {Waldmann}, \&
  {Tinetti}}]{Al-Refaie2021}
{Al-Refaie}, A.~F., {Changeat}, Q., {Waldmann}, I.~P., \& {Tinetti}, G. 2021,
  \apj, 917, 37, \dodoi{10.3847/1538-4357/ac0252}

\bibitem[{{Allard} {et~al.}(2003){Allard}, {Guillot}, {Ludwig}, {Hauschildt},
  {Schweitzer}, {Alexander}, \& {Ferguson}}]{Allard2003}
{Allard}, F., {Guillot}, T., {Ludwig}, H.-G., {et~al.} 2003, in Brown Dwarfs,
  ed. E.~{Mart{\'\i}n}, Vol. 211, 325

\bibitem[{{Barclay} {et~al.}(2021){Barclay}, {Kostov}, {Col{\'o}n}, {Quintana},
  {Schlieder}, {Louie}, {Gilbert}, \& {Mullally}}]{Barclay2021}
{Barclay}, T., {Kostov}, V.~B., {Col{\'o}n}, K.~D., {et~al.} 2021, \aj, 162,
  300, \dodoi{10.3847/1538-3881/ac2824}

\bibitem[{{Barstow} {et~al.}(2015){Barstow}, {Aigrain}, {Irwin}, {Kendrew}, \&
  {Fletcher}}]{Barstow2015}
{Barstow}, J.~K., {Aigrain}, S., {Irwin}, P.~G.~J., {Kendrew}, S., \&
  {Fletcher}, L.~N. 2015, \mnras, 448, 2546, \dodoi{10.1093/mnras/stv186}

\bibitem[{{Bean} {et~al.}(2010){Bean}, {Miller-Ricci Kempton}, \&
  {Homeier}}]{Bean2010}
{Bean}, J.~L., {Miller-Ricci Kempton}, E., \& {Homeier}, D. 2010, \nat, 468,
  669, \dodoi{10.1038/nature09596}

\bibitem[{Beichman {et~al.}(2014)Beichman, Benneke, Knutson, Smith, Lagage,
  Dressing, Latham, Lunine, Birkmann, Ferruit,
  {et~al.}}]{beichman2014observations}
Beichman, C., Benneke, B., Knutson, H., {et~al.} 2014, Publications of the
  Astronomical Society of the Pacific, 126, 1134

\bibitem[{{Benneke} {et~al.}(2019){Benneke}, {Wong}, {Piaulet}, {Knutson},
  {Lothringer}, {Morley}, {Crossfield}, {Gao}, {Greene}, {Dressing},
  {Dragomir}, {Howard}, {McCullough}, {Kempton}, {Fortney}, \&
  {Fraine}}]{Benneke2019b}
{Benneke}, B., {Wong}, I., {Piaulet}, C., {et~al.} 2019, \apjl, 887, L14,
  \dodoi{10.3847/2041-8213/ab59dc}

\bibitem[{{Buchner} {et~al.}(2014){Buchner}, {Georgakakis}, {Nandra}, {Hsu},
  {Rangel}, {Brightman}, {Merloni}, {Salvato}, {Donley}, \&
  {Kocevski}}]{Buchner2014}
{Buchner}, J., {Georgakakis}, A., {Nandra}, K., {et~al.} 2014, \aap, 564, A125,
  \dodoi{10.1051/0004-6361/201322971}

\bibitem[{{Claret}(2000)}]{Claret2000}
{Claret}, A. 2000, \aap, 363, 1081

\bibitem[{{Cloutier} {et~al.}(2019){Cloutier}, {Astudillo-Defru}, {Bonfils},
  {Jenkins}, {Berdi{\~n}as}, {Ricker}, {Vanderspek}, {Latham}, {Seager},
  {Winn}, {Jenkins}, {Almenara}, {Bouchy}, {Delfosse}, {D{\'\i}az},
  {D{\'\i}az}, {Doyon}, {Figueira}, {Forveille}, {Kurtovic}, {Lovis}, {Mayor},
  {Menou}, {Morgan}, {Morris}, {Muirhead}, {Murgas}, {Pepe}, {Santos},
  {S{\'e}gransan}, {Smith}, {Tenenbaum}, {Torres}, {Udry}, {Vezie}, \&
  {Villasenor}}]{Cloutier2019}
{Cloutier}, R., {Astudillo-Defru}, N., {Bonfils}, X., {et~al.} 2019, \aap, 629,
  A111, \dodoi{10.1051/0004-6361/201935957}

\bibitem[{{Damiano} {et~al.}(2017){Damiano}, {Morello}, {Tsiaras}, {Zingales},
  \& {Tinetti}}]{Damiano2017}
{Damiano}, M., {Morello}, G., {Tsiaras}, A., {Zingales}, T., \& {Tinetti}, G.
  2017, \aj, 154, 39, \dodoi{10.3847/1538-3881/aa738b}

\bibitem[{Dasgupta \& Grewal(2019)}]{dasgupta2019origin}
Dasgupta, R., \& Grewal, D.~S. 2019, Deep carbon, 4

\bibitem[{De~Wit {et~al.}(2018)De~Wit, Wakeford, Lewis, Delrez, Gillon, Selsis,
  Leconte, Demory, Bolmont, Bourrier, {et~al.}}]{de2018atmospheric}
De~Wit, J., Wakeford, H.~R., Lewis, N.~K., {et~al.} 2018, Nature Astronomy, 2,
  214

\bibitem[{{Demangeon} {et~al.}(2021){Demangeon}, {Zapatero Osorio}, {Alibert},
  {Barros}, {Adibekyan}, {Tabernero}, {Antoniadis-Karnavas}, {Camacho},
  {Su{\'a}rez Mascare{\~n}o}, {Oshagh}, {Micela}, {Sousa}, {Lovis}, {Pepe},
  {Rebolo}, {Cristiani}, {Santos}, {Allart}, {Allende Prieto}, {Bossini},
  {Bouchy}, {Cabral}, {Damasso}, {Di Marcantonio}, {D'Odorico}, {Ehrenreich},
  {Faria}, {Figueira}, {G{\'e}nova Santos}, {Haldemann}, {Hara}, {Gonz{\'a}lez
  Hern{\'a}ndez}, {Lavie}, {Lillo-Box}, {Lo Curto}, {Martins}, {M{\'e}gevand},
  {Mehner}, {Molaro}, {Nunes}, {Pall{\'e}}, {Pasquini}, {Poretti}, {Sozzetti},
  \& {Udry}}]{Damangeon2021}
{Demangeon}, O.~D.~S., {Zapatero Osorio}, M.~R., {Alibert}, Y., {et~al.} 2021,
  \aap, 653, A41, \dodoi{10.1051/0004-6361/202140728}

\bibitem[{{Deming} {et~al.}(2013){Deming}, {Wilkins}, {McCullough}, {Burrows},
  {Fortney}, {Agol}, {Dobbs-Dixon}, {Madhusudhan}, {Crouzet}, {Desert},
  {Gilliland}, {Haynes}, {Knutson}, {Line}, {Magic}, {Mandell}, {Ranjan},
  {Charbonneau}, {Clampin}, {Seager}, \& {Showman}}]{Deming2013}
{Deming}, D., {Wilkins}, A., {McCullough}, P., {et~al.} 2013, \apj, 774, 95,
  \dodoi{10.1088/0004-637X/774/2/95}

\bibitem[{Dong {et~al.}(2018)Dong, Jin, Lingam, Airapetian, Ma, \& van~der
  Holst}]{dong2018atmospheric}
Dong, C., Jin, M., Lingam, M., {et~al.} 2018, Proceedings of the National
  Academy of Sciences, 115, 260

\bibitem[{{Evans} {et~al.}(2016){Evans}, {Sing}, {Wakeford}, {Nikolov},
  {Ballester}, {Drummond}, {Kataria}, {Gibson}, {Amundsen}, \&
  {Spake}}]{Evans2016}
{Evans}, T.~M., {Sing}, D.~K., {Wakeford}, H.~R., {et~al.} 2016, \apj, 822, L4,
  \dodoi{10.3847/2041-8205/822/1/L4}

\bibitem[{Feroz \& Hobson(2007)}]{Feroz2007}
Feroz, F., \& Hobson, M.~P. 2007

\bibitem[{Feroz {et~al.}(2009)Feroz, Hobson, \& Bridges}]{Feroz2009}
Feroz, F., Hobson, M.~P., \& Bridges, M. T.~B. 2009

\bibitem[{Feroz {et~al.}(2013)Feroz, Hobson, Cameron, \& Pettitt}]{Feroz2013}
Feroz, F., Hobson, M.~P., Cameron, E., \& Pettitt, A.~N. 2013

\bibitem[{{Foreman-Mackey} {et~al.}(2013){Foreman-Mackey}, {Hogg}, {Lang}, \&
  {Goodman}}]{Foreman-Mackey2013}
{Foreman-Mackey}, D., {Hogg}, D.~W., {Lang}, D., \& {Goodman}, J. 2013, \pasp,
  125, 306, \dodoi{10.1086/670067}

\bibitem[{{Fraine} {et~al.}(2014){Fraine}, {Deming}, {Benneke}, {Knutson},
  {Jord{\'a}n}, {Espinoza}, {Madhusudhan}, {Wilkins}, \&
  {Todorov}}]{Fraine2014}
{Fraine}, J., {Deming}, D., {Benneke}, B., {et~al.} 2014, \nat, 513, 526,
  \dodoi{10.1038/nature13785}

\bibitem[{{Freedman} {et~al.}(2014){Freedman}, {Lustig-Yaeger}, {Fortney},
  {Lupu}, {Marley}, \& {Lodders}}]{Freedman2014}
{Freedman}, R.~S., {Lustig-Yaeger}, J., {Fortney}, J.~J., {et~al.} 2014, \apjs,
  214, 25, \dodoi{10.1088/0067-0049/214/2/25}

\bibitem[{{Freedman} {et~al.}(2008){Freedman}, {Marley}, \&
  {Lodders}}]{Freedman2008}
{Freedman}, R.~S., {Marley}, M.~S., \& {Lodders}, K. 2008, \apjs, 174, 504,
  \dodoi{10.1086/521793}

\bibitem[{Gaillard {et~al.}(2022)Gaillard, Bernadou, Roskosz, Bouhifd,
  Marrocchi, Iacono-Marziano, Moreira, Scaillet, \&
  Rogerie}]{gaillard2022redox}
Gaillard, F., Bernadou, F., Roskosz, M., {et~al.} 2022, Earth and Planetary
  Science Letters, 577, 117255

\bibitem[{{Haynes} {et~al.}(2015){Haynes}, {Mandell}, {Madhusudhan}, {Deming},
  \& {Knutson}}]{Haynes2015}
{Haynes}, K., {Mandell}, A.~M., {Madhusudhan}, N., {Deming}, D., \& {Knutson},
  H. 2015, \apj, 806, 146, \dodoi{10.1088/0004-637X/806/2/146}

\bibitem[{{Herrero} {et~al.}(2016){Herrero}, {Ribas}, {Jordi}, {Morales},
  {Perger}, \& {Rosich}}]{Herrero2016}
{Herrero}, E., {Ribas}, I., {Jordi}, C., {et~al.} 2016, \aap, 586, A131,
  \dodoi{10.1051/0004-6361/201425369}

\bibitem[{{Horne}(1986)}]{Horne1986}
{Horne}, K. 1986, Publications of the Astronomical Society of the Pacific, 98,
  609, \dodoi{10.1086/131801}

\bibitem[{Hu(2021)}]{hu2021photochemistry}
Hu, R. 2021, The Astrophysical Journal, 921, 27

\bibitem[{Hu {et~al.}(2012)Hu, Ehlmann, \& Seager}]{hu2012theoretical}
Hu, R., Ehlmann, B.~L., \& Seager, S. 2012, The Astrophysical Journal, 752, 7

\bibitem[{{Huitson} {et~al.}(2013){Huitson}, {Sing}, {Pont}, {Fortney},
  {Burrows}, {Wilson}, {Ballester}, {Nikolov}, {Gibson}, {Deming}, {Aigrain},
  {Evans}, {Henry}, {Lecavelier des Etangs}, {Showman}, {Vidal-Madjar}, \&
  {Zahnle}}]{Huitson2013}
{Huitson}, C.~M., {Sing}, D.~K., {Pont}, F., {et~al.} 2013, \mnras, 434, 3252,
  \dodoi{10.1093/mnras/stt1243}

\bibitem[{{Johnstone}(2020)}]{Johnstone2020}
{Johnstone}, C.~P. 2020, \apj, 890, 79, \dodoi{10.3847/1538-4357/ab6224}

\bibitem[{{Johnstone} {et~al.}(2021){Johnstone}, {Lammer}, {Kislyakova},
  {Scherf}, \& {G{\"u}del}}]{Johnstone2021}
{Johnstone}, C.~P., {Lammer}, H., {Kislyakova}, K.~G., {Scherf}, M., \&
  {G{\"u}del}, M. 2021, Earth and Planetary Science Letters, 576, 117197,
  \dodoi{10.1016/j.epsl.2021.117197}

\bibitem[{{Jord{\'a}n} {et~al.}(2013){Jord{\'a}n}, {Espinoza}, {Rabus},
  {Eyheramendy}, {Sing}, {D{\'e}sert}, {Bakos}, {Fortney}, {L{\'o}pez-Morales},
  {Maxted}, {Triaud}, \& {Szentgyorgyi}}]{Jordan2013}
{Jord{\'a}n}, A., {Espinoza}, N., {Rabus}, M., {et~al.} 2013, \apj, 778, 184,
  \dodoi{10.1088/0004-637X/778/2/184}

\bibitem[{{Kempton} {et~al.}(2017){Kempton}, {Lupu}, {Owusu-Asare}, {Slough},
  \& {Cale}}]{Kempton2017}
{Kempton}, E. M.~R., {Lupu}, R., {Owusu-Asare}, A., {Slough}, P., \& {Cale}, B.
  2017, \pasp, 129, 044402, \dodoi{10.1088/1538-3873/aa61ef}

\bibitem[{Kite \& Barnett(2020)}]{kite2020exoplanet}
Kite, E.~S., \& Barnett, M.~N. 2020, Proceedings of the National Academy of
  Sciences, 117, 18264

\bibitem[{Kite {et~al.}(2020)Kite, Fegley~Jr, Schaefer, \&
  Ford}]{kite2020atmosphere}
Kite, E.~S., Fegley~Jr, B., Schaefer, L., \& Ford, E.~B. 2020, The
  Astrophysical Journal, 891, 111

\bibitem[{{Kite} {et~al.}(2009){Kite}, {Manga}, \& {Gaidos}}]{Kite2009}
{Kite}, E.~S., {Manga}, M., \& {Gaidos}, E. 2009, \apj, 700, 1732,
  \dodoi{10.1088/0004-637X/700/2/1732}

\bibitem[{{Kite} \& {Schaefer}(2021)}]{Kite2021}
{Kite}, E.~S., \& {Schaefer}, L. 2021, \apjl, 909, L22,
  \dodoi{10.3847/2041-8213/abe7dc}

\bibitem[{{Knutson} {et~al.}(2014){Knutson}, {Benneke}, {Deming}, \&
  {Homeier}}]{Knutson2014}
{Knutson}, H.~A., {Benneke}, B., {Deming}, D., \& {Homeier}, D. 2014, \nat,
  505, 66, \dodoi{10.1038/nature12887}

\bibitem[{{Knutson} {et~al.}(2007){Knutson}, {Charbonneau}, {Noyes}, {Brown},
  \& {Gilliland}}]{Knutson2007}
{Knutson}, H.~A., {Charbonneau}, D., {Noyes}, R.~W., {Brown}, T.~M., \&
  {Gilliland}, R.~L. 2007, \apj, 655, 564, \dodoi{10.1086/510111}

\bibitem[{{Kostov} {et~al.}(2019){Kostov}, {Schlieder}, {Barclay}, {Quintana},
  {Col{\'o}n}, {Brande}, {Collins}, {Feinstein}, {Hadden}, {Kane}, {Kreidberg},
  {Kruse}, {Lam}, {Matthews}, {Montet}, {Pozuelos}, {Stassun}, {Winters},
  {Ricker}, {Vanderspek}, {Latham}, {Seager}, {Winn}, {Jenkins}, {Afanasev},
  {Armstrong}, {Arney}, {Boyd}, {Barentsen}, {Barkaoui}, {Batalha}, {Beichman},
  {Bayliss}, {Burke}, {Burdanov}, {Cacciapuoti}, {Carson}, {Charbonneau},
  {Christiansen}, {Ciardi}, {Clampin}, {Collins}, {Conti}, {Coughlin},
  {Covone}, {Crossfield}, {Delrez}, {Domagal-Goldman}, {Dressing}, {Ducrot},
  {Essack}, {Everett}, {Fauchez}, {Foreman-Mackey}, {Gan}, {Gilbert}, {Gillon},
  {Gonzales}, {Hamann}, {Hedges}, {Hocutt}, {Hoffman}, {Horch}, {Horne},
  {Howell}, {Hynes}, {Ireland}, {Irwin}, {Isopi}, {Jensen}, {Jehin},
  {Kaltenegger}, {Kielkopf}, {Kopparapu}, {Lewis}, {Lopez}, {Lissauer}, {Mann},
  {Mallia}, {Mandell}, {Matson}, {Mazeh}, {Monsue}, {Moran}, {Moran}, {Morley},
  {Morris}, {Muirhead}, {Mukai}, {Mullally}, {Mullally}, {Murray}, {Narita},
  {Palle}, {Pidhorodetska}, {Quinn}, {Relles}, {Rinehart}, {Ritsko},
  {Rodriguez}, {Rowden}, {Rowe}, {Sebastian}, {Sefako}, {Shahaf}, {Shporer},
  {Ta{\~n}{\'o}n Reyes}, {Tenenbaum}, {Ting}, {Twicken}, {van Belle}, {Vega},
  {Volosin}, {Walkowicz}, \& {Youngblood}}]{Kostov2019}
{Kostov}, V.~B., {Schlieder}, J.~E., {Barclay}, T., {et~al.} 2019, \aj, 158,
  32, \dodoi{10.3847/1538-3881/ab2459}

\bibitem[{{Kreidberg}(2015)}]{Kreidberg2015}
{Kreidberg}, L. 2015, \pasp, 127, 1161, \dodoi{10.1086/683602}

\bibitem[{{Kreidberg} {et~al.}(2014){Kreidberg}, {Bean}, {D{\'e}sert},
  {Benneke}, {Deming}, {Stevenson}, {Seager}, {Berta-Thompson}, {Seifahrt}, \&
  {Homeier}}]{Kreidberg2014}
{Kreidberg}, L., {Bean}, J.~L., {D{\'e}sert}, J.-M., {et~al.} 2014, \nat, 505,
  69, \dodoi{10.1038/nature12888}

\bibitem[{{Kreidberg} {et~al.}(2018){Kreidberg}, {Line}, {Parmentier},
  {Stevenson}, {Louden}, {Bonnefoy}, {Faherty}, {Henry}, {Williamson},
  {Stassun}, {Beatty}, {Bean}, {Fortney}, {Showman}, {D{\'e}sert}, \&
  {Arcangeli}}]{Kreidberg2018}
{Kreidberg}, L., {Line}, M.~R., {Parmentier}, V., {et~al.} 2018, \aj, 156, 17,
  \dodoi{10.3847/1538-3881/aac3df}

\bibitem[{Kreidberg {et~al.}(2019)Kreidberg, Koll, Morley, Hu, Schaefer,
  Deming, Stevenson, Dittmann, Vanderburg, Berardo,
  {et~al.}}]{kreidberg2019absence}
Kreidberg, L., Koll, D.~D., Morley, C., {et~al.} 2019, Nature, 573, 87

\bibitem[{Liggins {et~al.}(2020)Liggins, Shorttle, \& Rimmer}]{liggins2020can}
Liggins, P., Shorttle, O., \& Rimmer, P.~B. 2020, Earth and Planetary Science
  Letters, 550, 116546

\bibitem[{{Lupu} {et~al.}(2014){Lupu}, {Zahnle}, {Marley}, {Schaefer},
  {Fegley}, {Morley}, {Cahoy}, {Freedman}, \& {Fortney}}]{Lupu2014}
{Lupu}, R.~E., {Zahnle}, K., {Marley}, M.~S., {et~al.} 2014, \apj, 784, 27,
  \dodoi{10.1088/0004-637X/784/1/27}

\bibitem[{{McCullough} \& {MacKenty}(2012)}]{McCullough2012}
{McCullough}, P., \& {MacKenty}, J. 2012, {Considerations for using Spatial
  Scans with WFC3}, Space Telescope WFC Instrument Science Report

\bibitem[{{McCullough} {et~al.}(2014){McCullough}, {Crouzet}, {Deming}, \&
  {Madhusudhan}}]{McCullough2014}
{McCullough}, P.~R., {Crouzet}, N., {Deming}, D., \& {Madhusudhan}, N. 2014,
  \apj, 791, 55, \dodoi{10.1088/0004-637X/791/1/55}

\bibitem[{Miguel(2019)}]{miguel2019observability}
Miguel, Y. 2019, Monthly Notices of the Royal Astronomical Society, 482, 2893

\bibitem[{Misener \& Schlichting(2021)}]{misener2021cool}
Misener, W., \& Schlichting, H.~E. 2021, Monthly Notices of the Royal
  Astronomical Society, 503, 5658

\bibitem[{{Morello} {et~al.}(2020){Morello}, {Claret}, {Martin-Lagarde},
  {Cossou}, {Tsiaras}, \& {Lagage}}]{Morello2020}
{Morello}, G., {Claret}, A., {Martin-Lagarde}, M., {et~al.} 2020, \aj, 159, 75,
  \dodoi{10.3847/1538-3881/ab63dc}

\bibitem[{Mugnai {et~al.}(2021)Mugnai, Modirrousta-Galian, Edwards, Changeat,
  Bouwman, Morello, Al-Refaie, Baeyens, Bieger, Blain,
  {et~al.}}]{mugnai2021ares}
Mugnai, L.~V., Modirrousta-Galian, D., Edwards, B., {et~al.} 2021, The
  Astronomical Journal, 161, 284

\bibitem[{{Nikolov} {et~al.}(2015){Nikolov}, {Sing}, {Burrows}, {Fortney},
  {Henry}, {Pont}, {Ballester}, {Aigrain}, {Wilson}, {Huitson}, {Gibson},
  {D{\'e}sert}, {Lecavelier Des Etangs}, {Showman}, {Vidal-Madjar}, {Wakeford},
  \& {Zahnle}}]{Nikolov2015}
{Nikolov}, N., {Sing}, D.~K., {Burrows}, A.~S., {et~al.} 2015, \mnras, 447,
  463, \dodoi{10.1093/mnras/stu2433}

\bibitem[{{Pont} {et~al.}(2008){Pont}, {Knutson}, {Gilliland}, {Moutou}, \&
  {Charbonneau}}]{Pont2008}
{Pont}, F., {Knutson}, H., {Gilliland}, R.~L., {Moutou}, C., \& {Charbonneau},
  D. 2008, \mnras, 385, 109, \dodoi{10.1111/j.1365-2966.2008.12852.x}

\bibitem[{{Rackham} {et~al.}(2018){Rackham}, {Apai}, \&
  {Giampapa}}]{Rackham2018}
{Rackham}, B.~V., {Apai}, D., \& {Giampapa}, M.~S. 2018, \apj, 853, 122,
  \dodoi{10.3847/1538-4357/aaa08c}

\bibitem[{{Rackham} {et~al.}(2019){Rackham}, {Apai}, \&
  {Giampapa}}]{Rackham2019}
---. 2019, \aj, 157, 96, \dodoi{10.3847/1538-3881/aaf892}

\bibitem[{{Ricker} {et~al.}(2015){Ricker}, {Winn}, {Vanderspek}, {Latham},
  {Bakos}, {Bean}, {Berta-Thompson}, {Brown}, {Buchhave}, {Butler}, {Butler},
  {Chaplin}, {Charbonneau}, {Christensen-Dalsgaard}, {Clampin}, {Deming},
  {Doty}, {De Lee}, {Dressing}, {Dunham}, {Endl}, {Fressin}, {Ge}, {Henning},
  {Holman}, {Howard}, {Ida}, {Jenkins}, {Jernigan}, {Johnson}, {Kaltenegger},
  {Kawai}, {Kjeldsen}, {Laughlin}, {Levine}, {Lin}, {Lissauer}, {MacQueen},
  {Marcy}, {McCullough}, {Morton}, {Narita}, {Paegert}, {Palle}, {Pepe},
  {Pepper}, {Quirrenbach}, {Rinehart}, {Sasselov}, {Sato}, {Seager},
  {Sozzetti}, {Stassun}, {Sullivan}, {Szentgyorgyi}, {Torres}, {Udry}, \&
  {Villasenor}}]{Ricker2015}
{Ricker}, G.~R., {Winn}, J.~N., {Vanderspek}, R., {et~al.} 2015, Journal of
  Astronomical Telescopes, Instruments, and Systems, 1, 014003,
  \dodoi{10.1117/1.JATIS.1.1.014003}

\bibitem[{Schaefer \& Fegley(2017)}]{schaefer2017redox}
Schaefer, L., \& Fegley, B. 2017, The Astrophysical Journal, 843, 120

\bibitem[{{Sing} {et~al.}(2011){Sing}, {Pont}, {Aigrain}, {Charbonneau},
  {D{\'e}sert}, {Gibson}, {Gilliland}, {Hayek}, {Henry}, {Knutson}, {Lecavelier
  Des Etangs}, {Mazeh}, \& {Shporer}}]{Sing2011}
{Sing}, D.~K., {Pont}, F., {Aigrain}, S., {et~al.} 2011, \mnras, 416, 1443,
  \dodoi{10.1111/j.1365-2966.2011.19142.x}

\bibitem[{{Sing} {et~al.}(2016){Sing}, {Fortney}, {Nikolov}, {Wakeford},
  {Kataria}, {Evans}, {Aigrain}, {Ballester}, {Burrows}, {Deming},
  {D{\'e}sert}, {Gibson}, {Henry}, {Huitson}, {Knutson}, {Lecavelier Des
  Etangs}, {Pont}, {Showman}, {Vidal-Madjar}, {Williamson}, \&
  {Wilson}}]{Sing2016}
{Sing}, D.~K., {Fortney}, J.~J., {Nikolov}, N., {et~al.} 2016, \nat, 529, 59,
  \dodoi{10.1038/nature16068}

\bibitem[{{Sivia} \& {Skilling}(2006)}]{Sivia2006}
{Sivia}, D., \& {Skilling}, J. 2006, {Data Analysis A Bayesian Tutorial}
  (Oxford University Press)

\bibitem[{{Skilling}(2004)}]{Skilling2004}
{Skilling}, J. 2004, in American Institute of Physics Conference Series, ed.
  R.~{Fischer}, R.~{Preuss}, \& U.~V. {Toussaint}, Vol. 735, 395--405

\bibitem[{Skilling(2006)}]{Skilling2006}
Skilling, J. 2006, Bayesian Analysis, 1, 833

\bibitem[{{Swain} {et~al.}(2008){Swain}, {Vasisht}, \& {Tinetti}}]{Swain2008}
{Swain}, M.~R., {Vasisht}, G., \& {Tinetti}, G. 2008, \nat, 452, 329,
  \dodoi{10.1038/nature06823}

\bibitem[{{Swain} {et~al.}(2009){Swain}, {Tinetti}, {Vasisht}, {Deroo},
  {Griffith}, {Bouwman}, {Chen}, {Yung}, {Burrows}, {Brown}, {Matthews},
  {Rowe}, {Kuschnig}, \& {Angerhausen}}]{Swain2009}
{Swain}, M.~R., {Tinetti}, G., {Vasisht}, G., {et~al.} 2009, \apj, 704, 1616,
  \dodoi{10.1088/0004-637X/704/2/1616}

\bibitem[{Swain {et~al.}(2021)Swain, Estrela, Roudier, Sotin, Rimmer, Valio,
  West, Pearson, Huber-Feely, \& Zellem}]{swain2021detection}
Swain, M.~R., Estrela, R., Roudier, G.~M., {et~al.} 2021, The Astronomical
  Journal, 161, 213

\bibitem[{{Tian}(2009)}]{Tian2009}
{Tian}, F. 2009, \apj, 703, 905, \dodoi{10.1088/0004-637X/703/1/905}

\bibitem[{{Tsiaras} {et~al.}(2016{\natexlab{a}}){Tsiaras}, {Waldmann},
  {Rocchetto}, {Varley}, {Morello}, {Damiano}, \&
  {Tinetti}}]{Tsiaras2016B2016ApJ...832..202T}
{Tsiaras}, A., {Waldmann}, I.~P., {Rocchetto}, M., {et~al.} 2016{\natexlab{a}},
  \apj, 832, 202, \dodoi{10.3847/0004-637X/832/2/202}

\bibitem[{{Tsiaras} {et~al.}(2019){Tsiaras}, {Waldmann}, {Tinetti}, {Tennyson},
  \& {Yurchenko}}]{Tsiaras2019}
{Tsiaras}, A., {Waldmann}, I.~P., {Tinetti}, G., {Tennyson}, J., \&
  {Yurchenko}, S.~N. 2019, Nature Astronomy, 3, 1086,
  \dodoi{10.1038/s41550-019-0878-9}

\bibitem[{{Tsiaras} {et~al.}(2016{\natexlab{b}}){Tsiaras}, {Rocchetto},
  {Waldmann}, {Venot}, {Varley}, {Morello}, {Damiano}, {Tinetti}, {Barton},
  {Yurchenko}, \& {Tennyson}}]{Tsiaras2016B2016ApJ...820...99T}
{Tsiaras}, A., {Rocchetto}, M., {Waldmann}, I.~P., {et~al.} 2016{\natexlab{b}},
  \apj, 820, 99, \dodoi{10.3847/0004-637X/820/2/99}

\bibitem[{{Tsiaras} {et~al.}(2018){Tsiaras}, {Waldmann}, {Zingales},
  {Rocchetto}, {Morello}, {Damiano}, {Karpouzas}, {Tinetti}, {McKemmish},
  {Tennyson}, \& {Yurchenko}}]{Tsiaras2018}
{Tsiaras}, A., {Waldmann}, I.~P., {Zingales}, T., {et~al.} 2018, \aj, 155, 156,
  \dodoi{10.3847/1538-3881/aaaf75}

\bibitem[{Turbet {et~al.}(2020)Turbet, Bolmont, Ehrenreich, Gratier, Leconte,
  Selsis, Hara, \& Lovis}]{turbet2020revised}
Turbet, M., Bolmont, E., Ehrenreich, D., {et~al.} 2020, Astronomy \&
  Astrophysics, 638, A41

\bibitem[{{Zahnle} \& {Catling}(2017)}]{Zahnle2017}
{Zahnle}, K.~J., \& {Catling}, D.~C. 2017, \apj, 843, 122,
  \dodoi{10.3847/1538-4357/aa7846}

\bibitem[{{Zellem} {et~al.}(2017){Zellem}, {Swain}, {Roudier}, {Shkolnik},
  {Creech-Eakman}, {Ciardi}, {Line}, {Iyer}, {Bryden}, {Llama}, \&
  {Fahy}}]{Zellem2017}
{Zellem}, R.~T., {Swain}, M.~R., {Roudier}, G., {et~al.} 2017, \apj, 844, 27,
  \dodoi{10.3847/1538-4357/aa79f5}

\bibitem[{Zhang {et~al.}(2018)Zhang, Zhou, Rackham, \& Apai}]{zhang2018near}
Zhang, Z., Zhou, Y., Rackham, B.~V., \& Apai, D. 2018, The Astronomical
  Journal, 156, 178

\bibitem[{{Zhou} {et~al.}(2017){Zhou}, {Apai}, {Lew}, \&
  {Schneider}}]{Zhou2017}
{Zhou}, Y., {Apai}, D., {Lew}, B. W.~P., \& {Schneider}, G. 2017, \aj, 153,
  243, \dodoi{10.3847/1538-3881/aa6481}

\end{thebibliography}
	}
	
    \appendix
	
	\section{\edit1{Limb darkening coefficients}}\label{sec:limb}
	
	\edit1{We used the \cite{Claret2000} formulation for the limb darkening effect. We calculated the four coefficients per spectral bins by using the \exot\ python package. The coefficients are reported in Tab. \ref{tab:limb}}
	
	\begin{table*}[!h]
	    \center
		\caption{Spectral bins and limb darkening coefficients used in the Analysis A.}
		\label{tab:limb}
		\begin{tabular}{c c c c c}
			\hline \hline
			Spectral bins [nm] & a1 & a2 & a3 & a4 \\
			\hline
			1110.8 - 1141.6 & 1.28798093 & -1.05907762 & 0.51823029 & -0.0874053 \\
			1141.6 - 1170.9 & 1.28615323 & -1.08744839 & 0.54951993 & -0.09832113 \\
			1170.9 - 1198.8 & 1.29932409 & -1.153228   & 0.62233941 & -0.12612099 \\
            1198.8 - 1225.7 & 1.31403541 & -1.20359881 & 0.66336835 & -0.13702432 \\
            1225.7 - 1252.2 & 1.3216165  & -1.24503919 & 0.71220673 & -0.15662467 \\
            1252.2 - 1279.1 & 1.31087844 & -1.22725216 & 0.6846187  & -0.14247216 \\
            1279.1 - 1305.8 & 1.33205793 & -1.32351788 & 0.80287643 & -0.19154707 \\
            1305.8 - 1332.1 & 1.33327291 & -1.34653724 & 0.82282639 & -0.19646486 \\
            1332.1 - 1358.6 & 1.35830426 & -1.43403262 & 0.91975243 & -0.23435426 \\
            1358.6 - 1386.0 & 1.35014645 & -1.4334713  & 0.9180878  & -0.23234897 \\
            1386.0 - 1414.0 & 1.34613423 & -1.44645255 & 0.93259913 & -0.23707517 \\
            1414.0 - 1442.5 & 1.34866834 & -1.47233056 & 0.95865921 & -0.24580419 \\
            1442.5 - 1471.9 & 1.38009113 & -1.58464562 & 1.09298133 & -0.30172869 \\
            1471.9 - 1502.7 & 1.34297544 & -1.51871236 & 1.02049438 & -0.27172675 \\
            1502.7 - 1534.5 & 1.31575231 & -1.47873437 & 0.97919057 & -0.25523316 \\
            1534.5 - 1568.2 & 1.32973852 & -1.56395607 & 1.08417177 & -0.29880839 \\
            1568.2 - 1604.2 & 1.29595229 & -1.52289987 & 1.04537638 & -0.28498663 \\
            1604.2 - 1643.2 & 1.28150047 & -1.58351836 & 1.14303257 & -0.32898329 \\
			\hline\hline
		\end{tabular}
	\end{table*}
	\newpage
    
    \section{Single-visit transmission spectra}\label{sec:single_visit}
    
\begin{figure*}[!h]
		\centering
		\includegraphics[angle=0, scale=0.7]{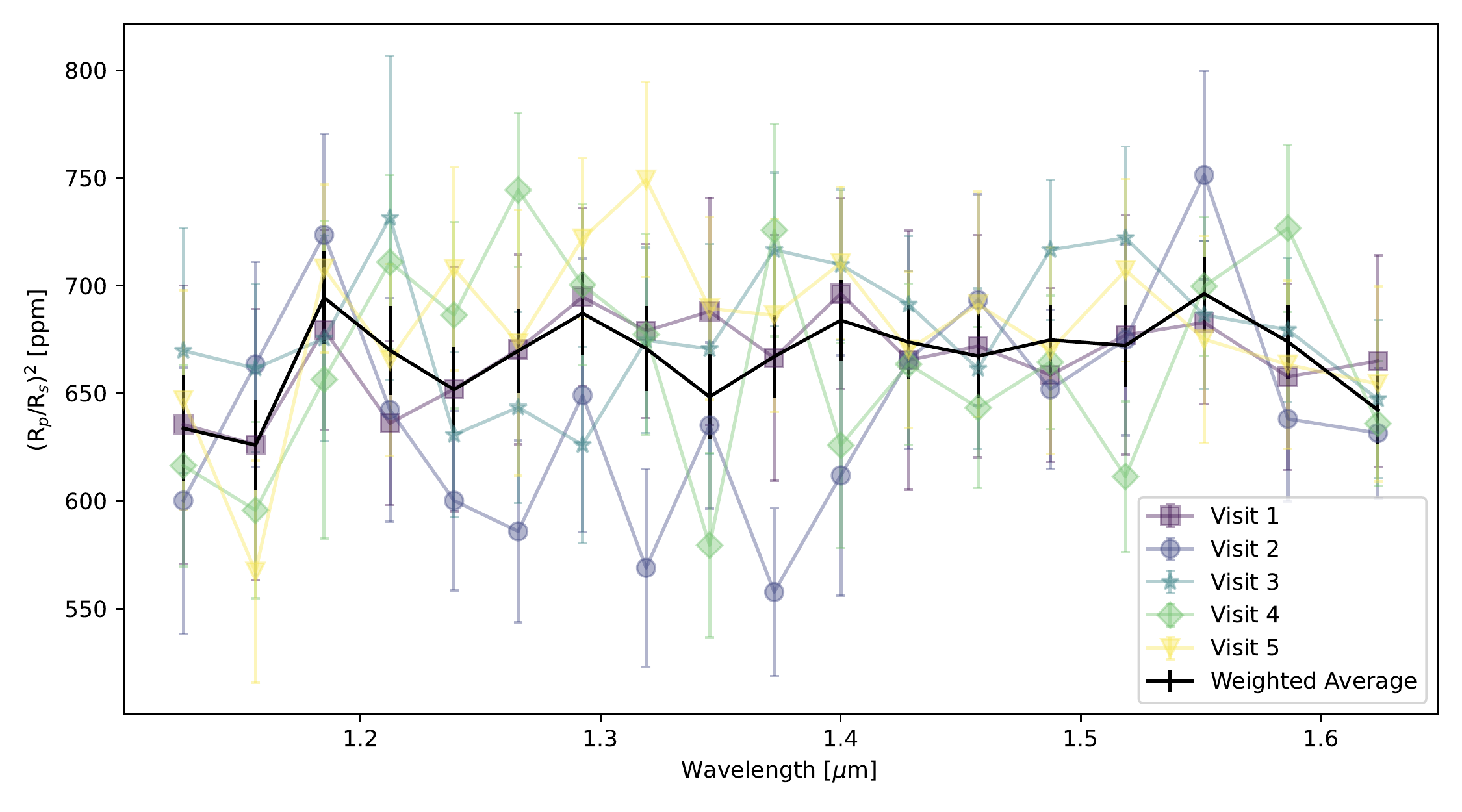}
		\caption{Colored : calculated 1D transmission spectrum of L98-59~b for each of the five HST visits resulting from the data analysis A described in Sec.~\ref{sec:model}. Black :  weighted average combined transmission spectrum. \label{fig:sp_single}}
	\end{figure*}

    Following the extraction and correction of the white light curves (see Sec. \ref{sec:model} and Sec. \ref{sec:result}), we fitted the spectral light curves to derive the 1D transmission spectrum. Fig.~\ref{fig:sp_single} shows the 1D transmission spectrum derived by using Analysis A for each of the five visits and and the combined weighted average. \edit1{In Tab. \ref{tab:diagnosis}, we report statistical diagnostics for each wavelength channel from the five visits. The standard deviation relative to the photon noise limit, $\overline{\sigma}$, is very close to unity.}
	
	\begin{table*}[!h]
	    \center
		\caption{\edit1{Spectral bins and residuals diagnostics for the each wavelength channel for the spectral light-curves processed using Analysis A. RMS: root mean squared of residuals; $\overline{\chi^2}$: reduced ${\chi^2}$; $\overline{\sigma}$: standard deviation relative to the photon noise limit; and R$_2$: auto-correlation. The diagnostics are presented as the mean and the standard deviation of the five visits.}}
		\label{tab:diagnosis}
		\begin{tabular}{c c c c c}
			\hline \hline
			Spectral bins [nm] & RMS [ppm] & $\overline{\chi^2}$ & $\overline{\sigma}$ & R$_2$\\
			\hline
			1110.8 - 1141.6 & 170 $\pm$ 16 & 1.062 $\pm$ 0.004 & 1.17 $\pm$ 0.11 & 0.21 $\pm$ 0.08\\
			1141.6 - 1170.9 & 155 $\pm$ 22 & 1.066 $\pm$ 0.008 & 1.09 $\pm$ 0.16 & 0.08 $\pm$ 0.04\\
			1170.9 - 1198.8 & 152 $\pm$ 13 & 1.066 $\pm$ 0.005 & 1.10 $\pm$ 0.09 & 0.16 $\pm$ 0.08\\
            1198.8 - 1225.7 & 170 $\pm$ 57 & 1.062 $\pm$ 0.004 & 1.25 $\pm$ 0.42 & 0.15 $\pm$ 0.05\\
            1225.7 - 1252.2 & 137 $\pm$ 12 & 1.066 $\pm$ 0.005 & 1.02 $\pm$ 0.09 & 0.13 $\pm$ 0.06\\
            1252.2 - 1279.1 & 147 $\pm$ 22 & 1.062 $\pm$ 0.004 & 1.11 $\pm$ 0.17 & 0.13 $\pm$ 0.07\\
            1279.1 - 1305.8 & 135 $\pm$ 6 & 1.064 $\pm$ 0.008 & 1.04 $\pm$ 0.04 & 0.12 $\pm$ 0.05\\
            1305.8 - 1332.1 & 136 $\pm$ 7 & 1.064 $\pm$ 0.005 & 1.05 $\pm$ 0.05 & 0.12 $\pm$ 0.08\\
            1332.1 - 1358.6 & 145 $\pm$ 9 & 1.064 $\pm$ 0.005 & 1.11 $\pm$ 0.08 & 0.10 $\pm$ 0.02\\
            1358.6 - 1386.0 & 152 $\pm$ 26 & 1.062 $\pm$ 0.004 & 1.17 $\pm$ 0.19 & 0.09 $\pm$ 0.05\\
            1386.0 - 1414.0 & 144 $\pm$ 19 & 1.064 $\pm$ 0.005 & 1.10 $\pm$ 0.15 & 0.10 $\pm$ 0.06\\
            1414.0 - 1442.5 & 137 $\pm$ 15 & 1.054 $\pm$ 0.012 & 1.05 $\pm$ 0.12 & 0.17 $\pm$ 0.08\\
            1442.5 - 1471.9 & 140 $\pm$ 10 & 1.054 $\pm$ 0.015 & 1.09 $\pm$ 0.08 & 0.12 $\pm$ 0.07\\
            1471.9 - 1502.7 & 126 $\pm$ 7 & 1.062 $\pm$ 0.004 & 1.00 $\pm$ 0.06 & 0.15 $\pm$ 0.06\\
            1502.7 - 1534.5 & 135 $\pm$ 6 & 1.062 $\pm$ 0.004 & 1.08 $\pm$ 0.05 & 0.13 $\pm$ 0.05\\
            1534.5 - 1568.2 & 137 $\pm$ 17 & 1.064 $\pm$ 0.005 & 1.11 $\pm$ 0.14 & 0.13 $\pm$ 0.07\\
            1568.2 - 1604.2 & 122 $\pm$ 8 & 1.062 $\pm$ 0.004 & 1.01 $\pm$ 0.06 & 0.14 $\pm$ 0.08\\
            1604.2 - 1643.2 & 132 $\pm$ 25 & 1.064 $\pm$ 0.005 & 1.11 $\pm$ 0.21 & 0.10 $\pm$ 0.06\\
			\hline\hline
		\end{tabular}
	\end{table*}
	\newpage
	
	\section{Posterior distributions}\label{sec:HCN}
	
\begin{figure*}[!h]
		\centering
		\includegraphics[angle=0, scale=0.29]{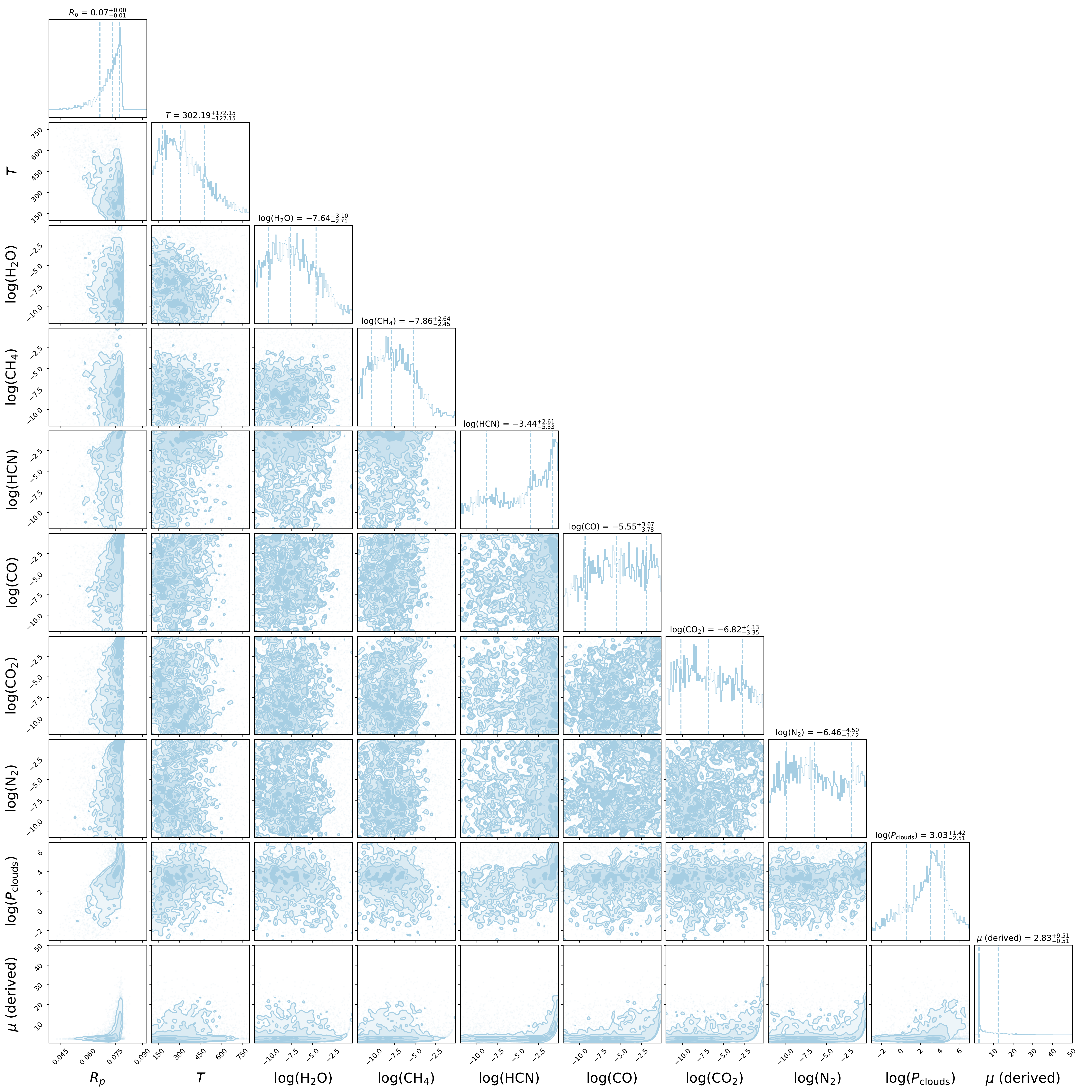}
		\caption{Posterior distributions of the retrieval on the 1D transmission spectrum of L98-59~b. The result suggests a light atmosphere (H$_2$-dominated) with clouds or haze and also hints the presence of HCN. The volume mixing ratio (VMR) of HCN has a distribution peak toward high values, but is otherwise not well constrained. All the other gases considered in the fitting show VMR distributions toward low values or completely flat, suggesting their absence in the atmosphere or a presence below the cloud deck not detectable by transmission spectroscopy. \edit1{The parameter $R_p$ in is defined as the radius at the bottom of the atmosphere, and therefore, when the atmosphere is more extended, the smaller this value must be to maintain consistency with the apparent radius of the planet.}\label{fig:HCN_post}}
	\end{figure*}

	The interpretation of the 1D spectrum shown in Fig.~\ref{fig:1d_final} has led to multiple scenarios that cannot be excluded. However, if the planet retains a light atmosphere, i.e. an H$_2$-dominated atmosphere, the statistical interpretation of the spectrum suggests that a significant amount of HCN might be present. We show the median-fit model from the spectral retrieval in Fig.~\ref{fig:best_fit} (orange line) and report the full posterior distributions here in Fig.~\ref{fig:HCN_post}.
	
\end{document}